\documentstyle[preprint,eqsecnum,aps]{revtex}
\tighten
\begin{document}
\preprint{UCONN-94-6}
\draft
\title{Maxwell-Chern-Simons theory in covariant and Coulomb gauges}
\author{Kurt Haller\thanks{khaller@uconnvm.uconn.edu} and Edwin
Lim-Lombridas\thanks{lombrida@uconnvm.uconn.edu}}
\address{Department of Physics, University of Connecticut, Storrs, Connecticut
06269}
\date{\today}
\maketitle
\begin{abstract}
We quantize Quantum Electrodynamics in $2+1$ dimensions coupled to a
Chern-Simons (CS) term and a charged spinor field, in covariant gauges and in
the Coulomb gauge. The resulting Maxwell-Chern-Simons (MCS) theory describes
charged fermions interacting with each other and with topologically massive
propagating photons. We impose Gauss's law and the gauge conditions and
investigate their effect on the dynamics and on the statistics of $n$-particle
states. We construct charged spinor states that obey Gauss's law and the gauge
conditions, and transform the theory to representations in which these states
constitute a Fock space. We demonstrate that, in these representations, the
nonlocal interactions between charges and between charges and transverse
currents, as well as the interactions between currents and massive propagating
photons, are identical in the different gauges we analyze in this and in
earlier work. We construct the generators of the Poincar\'e group, show that
they implement the Poincar\'e algebra, and explicitly demonstrate the effect of
rotations and Lorentz boosts on the particle states. We show that the
imposition of Gauss's law does not produce any ``exotic'' fractional
statistics. In the case of the covariant gauges, this demonstration makes use
of unitary transformations that provide charged particles with the gauge fields
required by Gauss's law, but that leave the anticommutator algebra of the
spinor fields untransformed. In the Coulomb gauge, we show that the
anticommutators of the spinor fields apply to the Dirac-Bergmann constraint
surfaces, on which Gauss's law and the gauge conditions obtain. We examine MCS
theory in the large CS coupling constant limit, and compare that limiting form
with CS theory, in which the Maxwell kinetic energy term is not included in the
Lagrangian.
\end{abstract}
\narrowtext

\section{Introduction}
In earlier work \cite{hl}, we discussed the quantization of $(2+1)$-dimensional
QED (QED$_3$) with a topological mass term---the so-called Maxwell-Chern-Simons
(MCS) theory---in the temporal $(A_0=0)$ gauge. In that work, we showed that
when Gauss's law is imposed on the particle excitations of a charged spinor
field, the charged particle states (fermions, in that case) do not develop
``exotic'' fractional statistics. We defined gauge-invariant fields that
create, from the vacuum, charged particle states that obey Gauss's law, and
nevertheless anticommute exactly like the gauge-dependent fields that create
the ``bare'' fermions from the vacuum. Moreover, the imposition of Gauss's law
does not cause the charged fermion states to acquire any arbitrary ``anyonic''
phases in a $2\pi$ rotation. These results contradict some widely accepted
conjectures about $(2+1)$-dimensional gauge theories with Chern-Simons (CS)
terms, in which exotic statistics and arbitrary rotational phases are regarded
as consequences of the imposition of Gauss's law on charged states
\cite{second}.

In this paper, we extend our investigation by studying the same model in the
covariant and the Coulomb gauges. Formulating this model in these different
gauges confirms our earlier results, and leads to new insights into its
gauge-independent observables and its particle states. First, we demonstrate
that the time-evolution operator for propagating particle states (i.e., the
Hamiltonian adjusted to comply with Gauss's law and the gauge condition) is
identical in the temporal, covariant, and Coulomb gauges, confirming that
identical predictions are obtained in every one of these gauges for all
questions that are in principle subject to empirical verification. This result
substantiates the consistency of our formulation of this model in the gauges we
have investigated. Our work in the Coulomb gauge, in which we apply the
Dirac-Bergmann (DB) procedure for imposing constraints \cite{dirac,bergmann},
supports our earlier demonstration that the implementation of Gauss's law does
not transform the statistical properties of the charged states of the spinor
field from the standard Fermi statistics to ``exotic'' fractional statistics.
We are able to corroborate this conclusion---already established in the
temporal gauge \cite{hl} and confirmed by an identical result for the covariant
gauges [Sec.~\ref{sec:anomalousrotation}]---by using the DB procedure in our
treatment of the Coulomb gauge to explicitly evaluate the anticommutator for
the spinor fields on the constraint surface on which all the theory's
constraints---including Gauss's law---apply. We use the covariant gauge
formulation of this model to obtain further insight into the kinematics of
$2\pi$ rotations for charged states in MCS theory, and to investigate the
effect of Lorentz boosts on the single propagating mode of the gauge field. One
element in this investigation is the demonstration that the operators used to
generate rotation and boosts implement the appropriate Poincar\'e algebra. In
this work, we also provide concrete illustrations of important abstract
principles---for example, illustrations of how operator-valued dynamical
variables develop gauge-independent forms in different gauges, even though they
are functionals of gauge and charged fermion fields whose forms and equations
of motion clearly reflect the choice of gauge. And finally, in this
investigation we explore the sense in which MCS theory approaches Chern-Simons
theory without a Maxwell kinetic energy term in the limit $m\rightarrow\infty$,
where $m$ is the topological mass.

\section{Formulation of MCS theory in covariant gauges}
\label{sec:formulation}
The Lagrangian for this model in the manifestly covariant gauges can be written
as
\begin{equation}
{\cal L} = -\case 1/4 F_{\mu\nu}F^{\mu\nu} + \case 1/4
m\epsilon_{\mu\nu\lambda}F^{\mu\nu}A^\lambda - G\partial_\mu A^\mu + \case 1/2
(1-\gamma)G^2 - j_\mu A^\mu + {\cal L}_{e\bar{e}}
\label{eq:calL1}
\end{equation}
where $F_{\mu\nu} = \partial_\mu A_\nu - \partial_\nu A_\mu$, $G$ is a
gauge-fixing field, and $\gamma$ is a parameter that permits ``tuning'' to
various alternative covariant gauges---for example, to the Feynman $(\gamma=0)$
and the Landau $(\gamma=1)$ gauges. ${\cal L}_{e\bar{e}}$ is the Lagrangian for
free fermions, and is given by
\begin{equation}
{\cal L}_{e\bar{e}} = \bar{\psi}(i\gamma_\mu\partial^\mu - M)\psi.
\end{equation}
$\psi$ is the two-component spinor field required for the $(2+1)$-dimensional
Dirac equation, and the three $\gamma^\mu$ are given in terms of the Pauli spin
matrices as $\gamma^0=-\sigma_3$, $\gamma^1=i\sigma_2$, and
$\gamma^2=-i\sigma_1$. The spinor currents take the form $j^\mu =
e\bar{\psi}\gamma^\mu\psi$.

Equation~(\ref{eq:calL1}) leads to the following Euler-Lagrange equations:
\begin{equation}
\partial^\nu F_{\mu\nu} - \case 1/2 m\epsilon_{\mu\nu\lambda}F^{\nu\lambda} -
\partial_\mu G + j_\mu = 0,
\label{eq:amperegauss}
\end{equation}
\begin{equation}
\partial_\mu A^\mu = (1-\gamma)G,
\label{eq:gaugefixing}
\end{equation}
and
\begin{equation}
(i\gamma_\mu D^\mu - M)\psi=0,
\end{equation}
where $D^\mu$ is the covariant derivative $D^\mu = \partial^\mu + ieA^\mu$.
Current conservation, $\partial_\mu j^\mu =0$, follows from
Eq.~(\ref{eq:amperegauss}) and implies that
\begin{equation}
\partial_\mu\partial^\mu G = 0.
\end{equation}

The gauge fields are not subject to any primary constraints in this formulation
of MCS theory in the covariant gauges, and all components of $A^\mu$ have
canonically conjugate momenta. These are defined by\footnote{For notational
simplicity, we will, from here on, generally use a noncovariant notation, in
which the subscript $l$ denotes a covariant component of a covariant quantity
(like $\partial_l$), a contravariant component of a contravariant quantity
(like $A_l$), or the contravariant component of the second rank tensor
$\Pi_l$.}
\begin{equation}
\Pi_l = F_{0l} + \case 1/2 m\epsilon_{ln}A_n,
\label{eq:Pil}
\end{equation}
for the momenta conjugate to the spatial components of the gauge field, $A_l$,
and by
\begin{equation}
\Pi_0 = -G,
\end{equation}
for the momentum conjugate to $A_0$. We are thus led to the Hamiltonian density
\begin{equation}
{\cal H} = {\cal H}_0 + {\cal H}_{\text{I}}
\label{eq:calH}
\end{equation}
where
\begin{eqnarray}
{\cal H}_0 &=& \case 1/2 \Pi_l\Pi_l + \case 1/4 F_{ln}F_{ln} -
\partial_lA_0\Pi_l + G\partial_lA_l + \case 1/8 m^2 A_lA_l\nonumber\\
&-& \case 1/2 (1-\gamma)G^2 - \case 1/4 m\epsilon_{ln}F_{ln}A_0 + \case 1/2
m\epsilon_{ln}A_l\Pi_n + {\cal H}_{e\bar{e}},
\end{eqnarray}
with ${\cal H}_{e\bar{e}} = \psi^\dagger(\gamma_0M -
i\gamma_0\gamma_l\partial_l)\psi$, and ${\cal H}_{\text{I}} = j_0A_0 - j_lA_l$.
The Hamiltonian $H$ is given by $H = \int d{\bf x}\ {\cal H}({\bf x})$ and can
be expressed as
\begin{equation}
H = H_0 + H_{\text{I}}
\end{equation}
where
\begin{eqnarray}
H_0 &=& \int d{\bf x}\ \left[\case 1/2 \Pi_l\Pi_l + \case 1/4 F_{ln}F_{ln} +
A_0(\partial_l\Pi_l - \case 1/4 m\epsilon_{ln}F_{ln}+j_0) +
G\partial_lA_l\right.\nonumber\\
&+& \left.\case 1/8 m^2 A_lA_l-j_lA_l-\case 1/2 (1-\gamma)G^2 + \case 1/2
m\epsilon_{ln}A_l\Pi_n\right] + H_{e\bar{e}},
\label{eq:Htotal}
\end{eqnarray}
after an integration by parts has been carried out; $H_{e\bar{e}} = \int d{\bf
x}\ {\cal H}_{e\bar{e}}({\bf x})$, and
\begin{equation}
H_{\text{I}} = \int d{\bf x}\ \left(j_0A_0 - j_lA_l\right).
\label{eq:HI}
\end{equation}

Since each gauge field has a canonical momentum, the equal-time commutation
(anticommutation) rules are canonical and are given by
\begin{equation}
[A_0({\bf x}),G({\bf y})] = -i\delta({\bf x-y}),
\label{eq:aog}
\end{equation}
\begin{equation}
[A_l({\bf x}),\Pi_n({\bf y})] = i\delta_{ln}\delta({\bf x-y}),
\label{eq:alpin}
\end{equation}
and
\begin{equation}
\{\psi({\bf x}),\psi^\dagger({\bf y})\} = \delta({\bf x-y}).
\end{equation}
In order to describe the particle states of this theory---in particular the
charged particle states that obey Gauss's law---we must represent the gauge and
spinor fields in terms of creation and annihilation operators for particle
excitations. In the case of the spinor fields in $2+1$ dimensions, a standard
representation that uses creation---$e^\dagger({\bf k})$ and
$\bar{e}^\dagger({\bf k})$---and annihilation---$e({\bf k})$ and $\bar{e}({\bf
k})$---operators for particle modes in definite momentum states is well known
\cite{hl}, and given by
\begin{equation}
\psi({\bf x}) = \sum_{\bf k}\sqrt{\frac{M}{\bar{\omega}_{k}}}\left[e({\bf
k})u({\bf k})e^{i{\bf k\cdot x}} + \bar{e}^\dagger({\bf k})v({\bf k})e^{-i{\bf
k\cdot x}}\right]
\label{eq:psi}
\end{equation}
and
\begin{equation}
\psi^\dagger({\bf x}) = \sum_{\bf
k}\sqrt{\frac{M}{\bar{\omega}_{k}}}\left[e^\dagger({\bf k})u^\dagger({\bf
k})e^{-i{\bf k\cdot x}} + \bar{e}({\bf k})v^\dagger({\bf k})e^{i{\bf k\cdot
x}}\right]
\end{equation}
where $\bar{\omega}_{k}=\sqrt{M^2+k^2}$; the excitation operators for the modes
of the spinor field obey the anticommutation rules $\{e({\bf k}),e^\dagger({\bf
q})\} = \delta_{\bf kq}$ and $\{\bar{e}({\bf k}),\bar{e}^\dagger({\bf q})\} =
\delta_{\bf kq}$ as well as $\{e({\bf k}),e({\bf q})\} = \{e^\dagger({\bf
k}),e^\dagger({\bf q})\} = 0$. The spinors $u({\bf k})$ and $v({\bf k})$ are
normalized so that $\bar{u}({\bf k})u({\bf k})=1$ and $\bar{v}({\bf k})v({\bf
k})=-1$ and are given by
\begin{equation}
u({\bf k}) =  \frac{1}{\sqrt{2M(M+\bar{\omega}_{k})}}\left(
\begin{array}{c}-k_{-}\\M+\bar{\omega}_{k}\end{array}\right)
\label{eq:ubfk}
\end{equation}
and
\begin{equation}
v({\bf k}) =  \frac{1}{\sqrt{2M(M+\bar{\omega}_{k})}}\left(
\begin{array}{c}M+\bar{\omega}_{k}\\ -k_{+}\end{array}\right)
\label{eq:vbfk}
\end{equation}
where $k_\pm = k_1 \pm ik_2$. In a later section of this paper, in which we
examine the phase changes effected by $2\pi$ rotations, we will find it
convenient to represent the spinor field using creation and annihilation
operators for excitations in definite angular momentum states. Such a
representation was previously used in Ref.~\cite{hl}.\footnote{From now on, we
will refer to the excitation modes of the spinor fields as electrons and
positrons, even though we are working in a $(2+1)$-dimensional space.}

For the gauge field, ``standard'' representations are not available. A suitable
representation of the gauge fields must be constructed for each gauge. One such
representation was constructed for the temporal gauge in Ref.~\cite{hl};
others, applicable to the covariant and Coulomb gauges, will be given in this
paper. To be suitable, a representation must be consistent with the equal-time
commutation rules given in Eqs.~(\ref{eq:aog}) and (\ref{eq:alpin}). There are
too many commutation rules for the gauge field in a covariant gauge to be
satisfied with only the single excitation mode of the topologically massive
``photon'' that exists in MCS theory. Further modes, in the form of ghost
excitations, are required. These ghost modes are identical to the ones that
appear in the temporal gauge \cite{hl}, and that also are required in
$(3+1)$-dimensional QED (QED$_4$) in covariant and axial (except for the
spatial axial) gauges \cite{el3}. The excitation operators for the massive
photon are the annihilation operator $a({\bf k})$ and its adjoint creation
operator $a^\dagger({\bf k})$, which obey the commutation rule $[a({\bf
k}),a^\dagger({\bf q})]=\delta_{{\bf kq}}$. Ghost excitation operators exist in
pairs; in this work, we will use the ghost annihilation operators $a_Q({\bf
k})$ and $a_R({\bf k})$ and their respective adjoint creation operators
$a_Q^\star({\bf k})$ and $a_R^\star({\bf k})$ in representations of the gauge
field. Ghost states have zero norm, but the single-particle ghost states
$a_Q^\star({\bf k})|0\rangle$ and $a_R^\star({\bf k})|0\rangle$ have a
nonvanishing inner product; similar nonvanishing inner products also arise for
$n$-particle states with equal numbers of $Q$ and $R$ ghosts. These properties
of the ghost states are implemented by the commutator algebra
\begin{equation}
[a_Q({\bf k}),a_R^\star({\bf q})] = [a_R({\bf k}),a_Q^\star({\bf q})] =
\delta_{\bf kq}
\end{equation}
and
\begin{equation}
[a_Q({\bf k}),a_Q^\star({\bf q})] = [a_R({\bf k}),a_R^\star({\bf q})] = 0,
\end{equation}
which, in turn, imply that the unit operator in the one-particle ghost (OPG)
sector is
\begin{equation}
1_{\text{OPG}} = \sum_{\bf k}\left[a_Q^\star({\bf k})|0\rangle\langle
0|a_R({\bf k}) +
a_R^\star({\bf k})|0\rangle\langle 0|a_Q({\bf k})\right],
\end{equation}
and that the obvious generalization of that form apply in $n$-particle sectors.
The ghost excitations enable us to satisfy {\em all} the equal-time commutation
relations (ETCR), Eqs.~(\ref{eq:aog}) and (\ref{eq:alpin}), even though the
gauge field has only a single mode that corresponds to a propagating particle
excitation that can, in principle, be detected, and that carries energy and
momentum.

There is another criterion that a representation must satisfy in order to be
suitable: The photon modes (propagating and ghost) must appear in the
Hamiltonian for free, noninteracting propagating photons and charged particles
in such a manner that dynamical time-evolution never propagates state vectors
into that part of Hilbert space in which inner products between the two
different types of ghost states drain probability from the sector of Hilbert
space spanned by observable particle states.

A representation of the gauge fields that satisfies these requirements in
covariant gauges is given by
\begin{eqnarray}
A_l({\bf x}) &=& \sum_{\bf k}\frac{8ik\epsilon_{ln}k_n}{m^{5/2}}\left[a_Q({\bf
k})e^{i{\bf k\cdot x}}-a_Q^\star({\bf k})e^{-i{\bf k\cdot x}}\right]\nonumber\\
&+& (1-\gamma)\sum_{\bf k}\frac{2k_l}{m^{3/2}}\left[a_Q({\bf k})e^{i{\bf k\cdot
x}}+a_Q^\star({\bf k})e^{-i{\bf k\cdot x}}\right]\nonumber\\
&-&\sum_{\bf k}\frac{4k^2k_l}{m^{7/2}}\left[a_Q({\bf k})e^{i{\bf k\cdot
x}}+a_Q^\star({\bf k})e^{-i{\bf k\cdot x}}\right]\nonumber\\
&+&\sum_{\bf k}\frac{m^{3/2}k_l}{16k^3}\left[a_R({\bf k})e^{i{\bf k\cdot
x}}+a_R^\star({\bf k})e^{-i{\bf k\cdot x}}\right]\nonumber\\
&-&\sum_{\bf k}\frac{\sqrt{\omega_k}k_l}{\sqrt{2}mk}\left[a({\bf k})e^{i{\bf
k\cdot x}}+a^\dagger({\bf k})e^{-i{\bf k\cdot x}}\right]\nonumber\\
&+&\sum_{\bf k}\frac{i\epsilon_{ln}k_n}{k\sqrt{2\omega_k}}\left[a({\bf
k})e^{i{\bf k\cdot x}}-a^\dagger({\bf k})e^{-i{\bf k\cdot x}}\right],
\label{eq:Ail}
\end{eqnarray}
\begin{eqnarray}
\Pi_l({\bf x}) &=& -\sum_{\bf k}\frac{4ikk_l}{m^{3/2}}\left[a_Q({\bf
k})e^{i{\bf k\cdot x}}-a_Q^\star({\bf k})e^{-i{\bf k\cdot x}}\right]\nonumber\\
&+&(1-\gamma)\sum_{\bf k}\frac{\epsilon_{ln}k_n}{\sqrt{m}}\left[a_Q({\bf
k})e^{i{\bf k\cdot x}}+a_Q^\star({\bf k})e^{-i{\bf k\cdot x}}\right]\nonumber\\
&+&\sum_{\bf k}\frac{6k^2\epsilon_{ln}k_n}{m^{5/2}}\left[a_Q({\bf k})e^{i{\bf
k\cdot x}}+a_Q^\star({\bf k})e^{-i{\bf k\cdot x}}\right]\nonumber\\
&+&\sum_{\bf k}\frac{m^{5/2}\epsilon_{ln}k_n}{32k^3}\left[a_R({\bf k})e^{i{\bf
k\cdot x}}+a_R^\star({\bf k})e^{-i{\bf k\cdot x}}\right]\nonumber\\
&+&\sum_{\bf k}\frac{imk_l}{2^{3/2}k\sqrt{\omega_k}}\left[a({\bf k})e^{i{\bf
k\cdot x}}-a^\dagger({\bf k})e^{-i{\bf k\cdot x}}\right]\nonumber\\
&+&\sum_{\bf k}\frac{\sqrt{\omega_k}\epsilon_{ln}k_n}{2^{3/2}k}\left[a({\bf
k})e^{i{\bf k\cdot x}}+a^\dagger({\bf k})e^{-i{\bf k\cdot x}}\right],
\label{eq:Pilmomentum}
\end{eqnarray}
\begin{eqnarray}
A_0({\bf x}) &=& -\sum_{\bf k}\frac{4k^3}{m^{7/2}}\left[a_Q({\bf k})e^{i{\bf
k\cdot x}}+a_Q^\star({\bf k})e^{-i{\bf k\cdot x}}\right]\nonumber\\
&-&(1-\gamma)\sum_{\bf k}\frac{2k}{m^{3/2}}\left[a_Q({\bf k})e^{i{\bf k\cdot
x}}+a_Q^\star({\bf k})e^{-i{\bf k\cdot x}}\right]\nonumber\\
&+&\sum_{\bf k}\frac{m^{3/2}}{16k^2}\left[a_R({\bf k})e^{i{\bf k\cdot
x}}+a_R^\star({\bf k})e^{-i{\bf k\cdot x}}\right]\nonumber\\
&-&\sum_{\bf k}\frac{k}{m\sqrt{2\omega_k}}\left[a({\bf k})e^{i{\bf k\cdot
x}}+a^\dagger({\bf k})e^{-i{\bf k\cdot x}}\right],
\end{eqnarray}
and
\begin{eqnarray}
G({\bf x}) = \sum_{\bf k}\frac{8ik^2}{m^{3/2}}\left[a_Q({\bf k})e^{i{\bf k\cdot
x}}-a_Q^\star({\bf k})e^{-i{\bf k\cdot x}}\right].
\label{eq:G}
\end{eqnarray}
where $\omega_k=\sqrt{m^2+k^2}$. The electric and magnetic fields then are
\begin{eqnarray}
E_l({\bf x}) &=& -\sum_{\bf k}\frac{imk_l}{k\sqrt{2\omega_k}}\left[a({\bf
k})e^{i{\bf k\cdot x}}-a^\dagger({\bf k})e^{-i{\bf k\cdot x}}\right]\nonumber\\
&-&\sum_{\bf k}\frac{\sqrt{\omega_k}\epsilon_{ln}k_n}{\sqrt{2}k}\left[a({\bf
k})e^{i{\bf k\cdot x}}+a^\dagger({\bf k})e^{-i{\bf k\cdot x}}\right]\nonumber\\
&-&\sum_{\bf k}\frac{8k^2\epsilon_{ln}k_n}{m^{5/2}}\left[a_Q({\bf k})e^{i{\bf
k\cdot x}}+a_Q^\star({\bf k})e^{-i{\bf k\cdot x}}\right]
\label{eq:Elfield}
\end{eqnarray}
and
\begin{eqnarray}
B({\bf x}) &=& \sum_{\bf k}\frac{k}{\sqrt{2\omega_k}}\left[a({\bf k})e^{i{\bf
k\cdot x}}+a^\dagger({\bf k})e^{-i{\bf k\cdot x}}\right]\nonumber\\
&+& \sum_{\bf k}\frac{8k^3}{m^{5/2}}\left[a_Q({\bf k})e^{i{\bf k\cdot
x}}+a_Q^\star({\bf k})e^{-i{\bf k\cdot x}}\right].
\label{eq:Bfield}
\end{eqnarray}
When Eqs.~(\ref{eq:Ail})--(\ref{eq:G}) are substituted into $H_0$, the
Hamiltonian for noninteracting gauge fields and charged spinor fields, we
obtain
\begin{eqnarray}
H_0 &=& \sum_{\bf k}\frac{\omega_k}{2}\left[a^\dagger({\bf k})a({\bf k}) +
a({\bf k})a^\dagger({\bf k})\right]\nonumber\\
&+& \sum_{\bf k}k\left[a_Q^\star({\bf k})a_R({\bf k}) + a_Q({\bf
k})a_R^\star({\bf k})\right]\nonumber\\
&-&(1-\gamma)\sum_{\bf k}\frac{64k^4}{m^3}\,a_Q^\star({\bf k})a_Q({\bf k}) +
H_{e\bar{e}}.
\label{eq:H0Covariant}
\end{eqnarray}

We now turn to consider the Fock space in which the operators in this theory
act, and in which $H_0$ time translates state vectors. We can construct a Fock
space, $\{|h\rangle\}$, suitable for this model, on the foundation of the
perturbative vacuum, $|0\rangle$, which is annihilated by all the annihilation
operators, $a({\bf k})$, $a_Q({\bf k})$ and $a_R({\bf k})$, as well as $e({\bf
k})$ and $\bar{e}({\bf k})$. This perturbative Fock space includes all
multiparticle states, $|N\rangle$, consisting of observable, propagating
particles, i.e. electrons, positrons, and photons, created when $e^\dagger({\bf
k})$, $\bar{e}^\dagger({\bf k})$, and $a^\dagger({\bf k})$, respectively act on
$|0\rangle$. All such states, $|N\rangle$, are eigenstates of $H_0$. States in
which one of the varieties of ghost creation operator acts on one of these
multiparticle states $|N\rangle$, e.g. $a_Q^\star({\bf k})|N\rangle$ or
$a_Q^\star({\bf k}_1)a_Q^\star({\bf k}_2)|N\rangle$, have zero norm; they have
no probability of being observed, and have vanishing expectation values of
energy, momentum, as well as all other observables. We will designate the
subspace of $\{|h\rangle\}$ that consists of all states $|N\rangle$, and of all
states in which a chain of  $a_Q^\star({\bf k})$ operators [but {\em no}
$a_R^\star({\bf k})$ operators] acts on $|N\rangle$, as $\{|n\rangle\}$. States
in which both varieties of ghosts appear simultaneously, such as
$a_Q^\star({\bf k}_1)a_R^\star({\bf k}_2)|N\rangle$, also are in the Fock space
$\{|h\rangle\}$, but not in $\{|n\rangle\}$; because these states have a
nonvanishing norm and contain ghosts, they are not probabilistically
interpretable. Their appearance in the course of time evolution signals a
defect in the theory. Since the states $|N\rangle$ constitute the set of states
in $\{|n\rangle\}$ from which all zero norm states (i.e., the ones with ghost
constituents) have been excised, we will sometimes speak of the set of
$|N\rangle$ as a quotient space of observable propagating states. The
time-evolution operator, $\exp\left(-iH_0t\right)$, which excludes the effect
of the interaction Hamiltonian, has the important property that if it acts on a
state vector $|n_i\rangle$ in $\{|n\rangle\}$, it can only propagate it within
$\{|n\rangle\}$. We observe that the only parts of $H_0$ that could possibly
cause a state vector to leave the subspace $\{|n\rangle\}$, are those that
contain either $a_R^\star({\bf k})$ or $a_R({\bf k})$ operators. The only part
of $H_0$ that has that feature contains the combination of operators $\Gamma =
a_R^\star({\bf k})a_Q({\bf k})+a_Q^\star({\bf k})a_R({\bf k})$.
When $a_R({\bf k})$ acts on a state vector $|n_i\rangle$, it either annihilates
the vacuum, or it annihilates one of the $a_Q^\star({\bf k})$ operators in
$\{|n\rangle\}$. In the latter case, $\Gamma$ replaces the annihilated
$a_Q^\star({\bf k})$ operator with an identical one. When $a_Q({\bf k})$ acts
on a state vector $|n_i\rangle$, it always annihilates it. It is therefore
impossible for $\Gamma$ to produce a state vector external to $\{|n\rangle\}$,
in which an $a_R^\star({\bf k})$ operator acts on $|n_i\rangle$. The only
effect of $\Gamma$ is to translate $|n_i\rangle$ states within $\{|n\rangle\}$.

Substitution of Eqs.~(\ref{eq:Ail})--(\ref{eq:G}) into the $H_{\text{I}}$ in
Eq.~(\ref{eq:HI}) leads to an expression in which all gauge-field excitations
appear, including creation and annihilation operators for both varieties of
ghosts:
\begin{eqnarray}
H_{\text{I}} &=& -\sum_{\bf k}\frac{4k^3}{m^{7/2}}\left[a_Q({\bf k})j_0(-{\bf
k}) + a_Q^\star({\bf k})j_0({\bf k})\right]\nonumber\\
&-&(1-\gamma)\sum_{\bf k}\frac{2k}{m^{3/2}}\left[a_Q({\bf k})j_0(-{\bf k}) +
a_Q^\star({\bf k})j_0({\bf k})\right]\nonumber\\
&+&\sum_{\bf k}\frac{m^{3/2}}{16k^2}\left[a_R({\bf k})j_0(-{\bf k}) +
a_R^\star({\bf k})j_0({\bf k})\right]\nonumber\\
&-&\sum_{\bf k}\frac{k}{m\sqrt{2\omega_k}}\left[a({\bf k})j_0(-{\bf k}) +
a^\dagger({\bf k})j_0({\bf k})\right]\nonumber\\
&-&\sum_{\bf k}\frac{8ik\epsilon_{ln}k_n}{m^{5/2}}\left[a_Q({\bf k})j_l(-{\bf
k}) - a_Q^\star({\bf k})j_l({\bf k})\right]\nonumber\\
&-&(1-\gamma)\sum_{\bf k}\frac{2k_l}{m^{3/2}}\left[a_Q({\bf k})j_l(-{\bf k}) +
a_Q^\star({\bf k})j_l({\bf k})\right]\nonumber\\
&+&\sum_{\bf k}\frac{4k^2k_l}{m^{7/2}}\left[a_Q({\bf k})j_l(-{\bf k}) +
a_Q^\star({\bf k})j_l({\bf k})\right]\nonumber\\
&-&\sum_{\bf k}\frac{m^{3/2}k_l}{16k^3}\left[a_R({\bf k})j_l(-{\bf k}) +
a_R^\star({\bf k})j_l({\bf k})\right]\nonumber\\
&+&\sum_{\bf k}\frac{\sqrt{\omega_k}k_l}{\sqrt{2}mk}\left[a({\bf k})j_l(-{\bf
k}) + a^\dagger({\bf k})j_l({\bf k})\right]\nonumber\\
&-&\sum_{\bf k}\frac{i\epsilon_{ln}k_n}{k\sqrt{2\omega_k}}\left[a({\bf
k})j_l(-{\bf k}) - a^\dagger({\bf k})j_l({\bf k})\right].
\label{eq:Hinteraction}
\end{eqnarray}
$H_{\text{I}}$ contains terms with creation and annihilation operators for both
varieties of ghosts---$a_Q({\bf k})$ and $a_R({\bf k})$ as well as
$a_Q^\star({\bf k})$ and $a_R^\star({\bf k})$---and therefore threatens to
drive state vectors out of the subspace $\{|n\rangle\}$. The reason for this
apparent failure to maintain consistency is that Gauss's law and the gauge
condition $\partial_\mu A^\mu = 0$ have not yet been implemented. In the next
section, we will show how implementation of the constraints prevents the
catastrophic appearance of state vectors in which both varieties of ghosts
coincide.

\section{Implementation of Gauss's law and gauge condition}
\label{sec:implofgausslaw}
It is easily seen in Eq.~(\ref{eq:amperegauss}) that Gauss's law is not an
equation of motion in this theory. The operator ${\cal G}$ that is used to
implement Gauss's law in this model is
\begin{equation}
{\cal G} = -\case 1/2 m\epsilon_{ln}F_{ln} + \partial_lF_{0l} + j_0;
\end{equation}
Eq.~(\ref{eq:amperegauss}) includes an equation of motion that incorporates
${\cal G}$, in the form ${\cal G}=\partial_0G$, in much the same way as in the
temporal gauge \cite{hl}. Gauss's law, ${\cal G}=0$, still remains to be
implemented. Using Eq.~(\ref{eq:Pil}), ${\cal G}$ can also be represented as
\begin{equation}
{\cal G} = \partial_l\Pi_l - \case 1/4 m\epsilon_{ln}F_{ln} + j_0.
\label{eq:calGmomentum}
\end{equation}
Substitution of Eqs.~(\ref{eq:Ail}) and (\ref{eq:Pilmomentum}) into
Eq.~(\ref{eq:calGmomentum}) leads to
\begin{equation}
{\cal G}({\bf x}) = \sum_{\bf k}\frac{8k^3}{m^{3/2}}\left[a_Q({\bf k})e^{i{\bf
k\cdot x}}+a_Q^\star({\bf k})e^{-i{\bf k\cdot x}}\right]+j_0({\bf x}).
\end{equation}
We can write this as
\begin{equation}
{\cal G}({\bf x}) = \sum_{\bf k}\frac{8k^3}{m^{3/2}}\left[\Omega({\bf
k})e^{i{\bf k\cdot x}}+\Omega^\star({\bf k})e^{-i{\bf k\cdot x}}\right]
\label{eq:gaussomega}
\end{equation}
where $\Omega({\bf k})$ is defined by
\begin{equation}
\Omega({\bf k}) = a_Q({\bf k}) + \frac{m^{3/2}}{16k^3}\,j_0({\bf k});
\end{equation}
with $j_0({\bf k})=\int d{\bf x}\ j_0({\bf x})e^{-i{\bf k\cdot x}}$. Similarly,
we can express $G$ as
\begin{equation}
G({\bf x}) = \sum_{\bf k}\frac{8ik^2}{m^{3/2}}\left[\Omega({\bf k})e^{i{\bf
k\cdot x}}-\Omega^\star({\bf k})e^{-i{\bf k\cdot x}}\right].
\label{eq:gaugeomega}
\end{equation}
We can therefore implement Gauss's law and the gauge condition by embedding the
theory in a subspace $\{|\nu\rangle\}$ of another Fock space, in which all the
state vectors $|\nu\rangle$ satisfy the condition
\begin{equation}
\Omega({\bf k})|\nu\rangle = 0.
\label{eq:omegaspace}
\end{equation}
For all state vectors $|\nu\rangle$ and $|\nu^\prime\rangle$ in the physical
subspace $\{|\nu\rangle\}$, it can be seen from Eqs.~(\ref{eq:gaussomega}) and
(\ref{eq:gaugeomega}) that $\langle\nu^\prime|{\cal G}|\nu\rangle=0$ and
$\langle\nu^\prime|G|\nu\rangle=0$. Moreover, the condition $\Omega({\bf
k})|\nu\rangle = 0$, once established, continues to hold at all other times
because
\begin{equation}
[H,\Omega({\bf k})] = -k\Omega({\bf k})
\label{eq:HOmega}
\end{equation}
so that $e^{iHt}\Omega({\bf k})e^{-iHt}=\Omega({\bf k})e^{-ikt}$, and
$\Omega({\bf k})e^{-iHt}|\nu\rangle = 0$ follows from
Eq.~(\ref{eq:omegaspace}). This means that a state vector initially in the
physical subspace $\{|\nu\rangle\}$ will always remain entirely contained in it
as it develops under time evolution. These considerations show that the
subspace $\{|\nu\rangle\}$ must be used to secure the implementation of Gauss's
law and the gauge condition. We therefore direct our attention to this subspace
and note that $\{|\nu\rangle\}$ can be related unitarily to the subspace
$\{|n\rangle\}$ by the unitary transformation $U=e^{D}$ for which
\begin{equation}
U^{-1}\Omega({\bf k})U = a_Q({\bf k}),
\end{equation}
where $D$ is given by
\begin{eqnarray}
D &=& i\int d{\bf x}\,d{\bf y}\ \xi(|{\bf
x-y}|)\left[\epsilon_{ln}\partial_l\Pi_n({\bf x})-\case 1/2 m\partial_lA_l({\bf
x})\right]j_0({\bf y})\nonumber\\
&+&i\int d{\bf x}\,d{\bf y}\ \chi(|{\bf x-y}|)G({\bf x})j_0({\bf y})+i\int
d{\bf x}\,d{\bf y}\ \zeta(|{\bf x-y}|)G({\bf x})j_0({\bf y})\nonumber\\
&+& i\int d{\bf x}\,d{\bf y}\ \eta(|{\bf x-y}|)\left[\partial_l\Pi_l({\bf
x})+\case 1/2 m\epsilon_{ln}\partial_lA_n({\bf x})\right]j_0({\bf y})
\label{eq:Dconfig}
\end{eqnarray}
and where
\begin{equation}
\xi(|{\bf x-y}|) = \sum_{\bf k}\frac{1}{m}\,\frac{e^{i{\bf k\cdot(x-y)}}}{k^2},
\label{eq:xi}
\end{equation}
and
\begin{equation}
\chi(|{\bf x-y}|) = \sum_{\bf
k}\frac{1}{4}\left[(1-\gamma)+\frac{2k^2}{m^2}\right]\frac{e^{i{\bf
k\cdot(x-y)}}}{k^2}.
\label{eq:barxi}
\end{equation}
$\zeta(|{\bf x-y}|)$ and $\eta(|{\bf x-y}|)$ are two additional functions that
can be included. They can be expressed as
\begin{equation}
\zeta(|{\bf x-y}|) = -\sum_{\bf k}\frac{m^{3/2}\theta({\bf
k})}{8}\,\frac{e^{i{\bf k\cdot(x-y)}}}{k^2},
\end{equation}
and
\begin{equation}
\eta(|{\bf x-y}|) = \sum_{\bf k}\frac{m^{3/2}\phi({\bf k})}{8}\,\frac{e^{i{\bf
k\cdot(x-y)}}}{k^3}
\label{eq:eta}
\end{equation}
with $\theta({\bf k})$ and $\phi({\bf k})$ as some arbitrary real and even
functions of ${\bf k}$. The operator $D$ can also be expressed as
\begin{eqnarray}
D &=& \sum_{\bf k}\frac{m^{3/2}}{16k^3}\left[a_R({\bf k})j_0(-{\bf
k})-a_R^\star({\bf k})j_0({\bf k})\right]\nonumber\\
&+& \sum_{\bf k} \theta({\bf k})\left[a_Q({\bf k})j_0(-{\bf k})-a_Q^\star({\bf
k})j_0({\bf k})\right]\nonumber\\
&+& \sum_{\bf k} i\phi({\bf k})\left[a_Q({\bf k})j_0(-{\bf k})+a_Q^\star({\bf
k})j_0({\bf k})\right].
\label{eq:DQ}
\end{eqnarray}
The unitary operator $U$ can be used to establish a mapping that maps
$\Omega({\bf k}) \rightarrow a_Q({\bf k})$ and $\{|\nu\rangle\} \rightarrow
\{|n\rangle\}$, where $\{|n\rangle\}$ is the subspace described in the
preceding section. The required state vectors $|n\rangle=U^{-1}|\nu\rangle$ are
those in which $a_Q^\star({\bf k})$ and $a^\dagger({\bf k})$, as well as
electron and positron creation operators, act on the perturbative vacuum state
$|0\rangle$. But no $a_R^\star({\bf k})$ operators may appear in the states
$|n\rangle$ for which $|\nu\rangle=U|n\rangle$ comprise the subspace
$\{|\nu\rangle\}$, since for states $|h_{R}\rangle$ in which $a_R^\star({\bf
k})$ operators act on the observable multiparticle state $|N\rangle$, $a_Q({\bf
k})|h_{R}\rangle \neq 0$. Therefore, $\Omega({\bf k})|\rho\rangle \neq 0$ for
the states $|\rho\rangle = U|h_{R}\rangle$.

We can exploit the existence of the unitary operator $U$ to systematically
construct the subspace $\{|\nu\rangle\}$ from $\{|n\rangle\}$. Alternatively,
we can use $U$ to transform all the operators we have previously defined, and
to use $\{|n\rangle\}$ as the representation of the subspace that fully
implements Gauss's law and the gauge condition with all interactions included
in the Hamiltonian. In order to make this latter choice, we note that, in the
mapping effected by $U$, operators ${\cal P}$ map into $\tilde{\cal P}$, i.e.,
$U^{-1}{\cal P}U = \tilde{\cal P}$. For example, $\tilde{\Omega}({\bf k}) =
a_Q({\bf k})$ so that Eq.~(\ref{eq:omegaspace}) takes the form $a_Q({\bf
k})|n\rangle=0$; and it is this equation that implements Gauss's law and the
gauge condition, $\partial_\mu A_\mu=0$, in this alternate transformed
representation. Using the Baker-Hausdorff-Campbell formula, we find that the
transformed Hamiltonian, $\tilde{H} =  U^{-1}HU$ is given by
\begin{equation}
\tilde{H} = H_0 + \tilde{H}_{\text{I}};
\end{equation}
here $H_0$ is the untransformed noninteracting Hamiltonian given in
Eq.~(\ref{eq:H0Covariant}) and
\begin{eqnarray}
\tilde{H}_{\text{I}} &=& \sum_{\bf k}\frac{i\epsilon_{ln}k_l}{mk^2}\,j_n({\bf
k})j_0(-{\bf k}) + \sum_{\bf k}\frac{1}{2m^2}\,j_0({\bf k})j_0(-{\bf
k})\nonumber\\
&+& \sum_{\bf k}\frac{3im^{3/2}k_l\phi({\bf k})}{16k^3}\,j_l({\bf k})j_0(-{\bf
k})\nonumber\\
&-&\sum_{\bf k}\frac{k}{m\sqrt{2\omega_k}}\left[a({\bf k})j_0(-{\bf k}) +
a^\dagger({\bf k})j_0({\bf k})\right]\nonumber\\
&+&\sum_{\bf k}\frac{\sqrt{\omega_k}k_l}{\sqrt{2}mk}\left[a({\bf k})j_l(-{\bf
k}) + a^\dagger({\bf k})j_l({\bf k})\right]\nonumber\\
&-&\sum_{\bf k}\frac{i\epsilon_{ln}k_n}{k\sqrt{2\omega_k}}\left[a({\bf
k})j_l(-{\bf k}) - a^\dagger({\bf k})j_l({\bf k})\right]\nonumber\\
&-&\sum_{\bf k}\frac{4k^3}{m^{7/2}}\left[a_Q({\bf k})j_0(-{\bf k}) +
a_Q^\star({\bf k})j_0({\bf k})\right]\nonumber\\
&-&\sum_{\bf k}\frac{8ik\epsilon_{ln}k_n}{m^{5/2}}\left[a_Q({\bf k})j_l(-{\bf
k}) - a_Q^\star({\bf k})j_l({\bf k})\right]\nonumber\\
&+&\sum_{\bf k}\frac{4k^2k_l}{m^{7/2}}\left[a_Q({\bf k})j_l(-{\bf k}) +
a_Q^\star({\bf k})j_l({\bf k})\right]\nonumber\\
&+&(1-\gamma)\sum_{\bf k}\frac{2k}{m^{3/2}}\left[a_Q({\bf k})j_0(-{\bf k}) +
a_Q^\star({\bf k})j_0({\bf k})\right]\nonumber\\
&-&(1-\gamma)\sum_{\bf k}\frac{2k_l}{m^{3/2}}\left[a_Q({\bf k})j_l(-{\bf k}) +
a_Q^\star({\bf k})j_l({\bf k})\right]\nonumber\\
&+&\sum_{\bf k}k\theta({\bf k})\left[a_Q({\bf k})j_0(-{\bf k}) + a_Q^\star({\bf
k})j_0({\bf k})\right]\nonumber\\
&-&\sum_{\bf k}ik\phi({\bf k})\left[a_Q({\bf k})j_0(-{\bf k}) - a_Q^\star({\bf
k})j_0({\bf k})\right]\nonumber\\
&+&\sum_{\bf k}k_l\theta({\bf k})\left[a_Q({\bf k})j_l(-{\bf k}) +
a_Q^\star({\bf k})j_l({\bf k})\right]\nonumber\\
&+&\sum_{\bf k}ik_l\phi({\bf k})\left[a_Q({\bf k})j_l(-{\bf k}) -
a_Q^\star({\bf k})j_l({\bf k})\right].
\label{eq:tildeHI}
\end{eqnarray}
The similarly transformed fields are
\begin{equation}
\tilde{A}_l({\bf x}) = A_l({\bf x}) - \sum_{\bf
k}\frac{i\epsilon_{ln}k_n}{mk^2}\,j_0({\bf k})e^{i{\bf k\cdot x}} + \sum_{\bf
k}\frac{im^{3/2}k_l\phi({\bf k})}{8k^3}\,j_0({\bf k})e^{i{\bf k\cdot x}},
\end{equation}
\begin{equation}
\tilde{\Pi}_l({\bf x}) = \Pi_l({\bf x}) + \sum_{\bf
k}\frac{ik_l}{2k^2}\,j_0({\bf k})e^{i{\bf k\cdot x}} + \sum_{\bf
k}\frac{im^{5/2}\epsilon_{ln}k_n\phi({\bf k})}{16k^3}\,j_0({\bf k})e^{i{\bf
k\cdot x}},
\end{equation}
\begin{eqnarray}
\tilde{A}_0({\bf x}) &=& A_0({\bf x}) + (1-\gamma)\sum_{\bf k}
\frac{1}{4k^2}\,j_0({\bf k})e^{i{\bf k\cdot x}}\nonumber\\
&+& \sum_{\bf k}\frac{1}{2m^2}\,j_0({\bf k})e^{i{\bf k\cdot x}} - \sum_{\bf
k}\frac{m^{3/2}\theta({\bf k})}{8k^2}\,j_0({\bf k})e^{i{\bf k\cdot x}},
\end{eqnarray}
\begin{equation}
\tilde{G}({\bf x}) = G({\bf x}) - \sum_{\bf k} \frac{i}{k}\,j_0({\bf
k})e^{i{\bf k\cdot x}},
\end{equation}
\begin{equation}
\tilde{\cal G} = \partial_l\Pi_l - \case 1/4 m\epsilon_{ln}F_{ln},
\end{equation}
and
\begin{equation}
\tilde{\psi}({\bf x}) = e^{{\cal D}({\bf x})}\psi({\bf x})
\label{eq:tildepsi}
\end{equation}
where
\begin{eqnarray}
{\cal D}({\bf x}) &=& -ie\int d{\bf y}\ \left\{\xi(|{\bf
x-y}|)\left[\epsilon_{ln}\partial_l\Pi_n({\bf y})-\case 1/2 m\partial_lA_l({\bf
y})\right] + \chi(|{\bf x-y}|)G({\bf y})\right.\nonumber\\
&+&\left.\eta(|{\bf x-y}|)\left[\partial_l\Pi_l({\bf x})+\case 1/2
m\epsilon_{ln}\partial_lA_n({\bf x})\right] + \zeta(|{\bf x-y}|)G({\bf
x})j_0({\bf y})\right\}.
\label{eq:calDpsi}
\end{eqnarray}
We note that under the gauge transformation $A_\mu \rightarrow A_\mu +
\partial_\mu\chi$, ${\cal D}({\bf x}) \rightarrow {\cal D}({\bf x}) - ie\chi$,
so that, since $\psi \rightarrow \psi\exp(ie\chi)$, $\tilde{\psi}$ is
gauge-invariant. The transformed electric and magnetic fields are
\begin{equation}
\tilde{E}_l({\bf x}) = E_l({\bf x})
\end{equation}
and
\begin{equation}
\tilde{B}({\bf x}) = B({\bf x}) + {\cal B}({\bf x})
\end{equation}
where $E_l({\bf x})$ and $B({\bf x})$ are given by Eqs.~(\ref{eq:Elfield}) and
(\ref{eq:Bfield}), respectively, and
\begin{equation}
{\cal B}({\bf x}) = -\frac{j_0({\bf x})}{m}.
\end{equation}
In this equivalent, alternative representation, $\exp\left(-i\tilde{H}t\right)$
is the time-translation operator. The time-translation operator will time
translate state vectors entirely within the physical subspace in the
transformed representation if $\tilde{H}$ is entirely devoid of $a_R^\star({\bf
k})$ and $a_R({\bf k})$ operators, or if it contains them at most in the
combination $\Gamma = a_R^\star({\bf k})a_Q({\bf k})+a_Q^\star({\bf k})a_R({\bf
k})$. Inspection of Eqs.~(\ref{eq:H0Covariant}) and (\ref{eq:tildeHI}) confirms
that $\tilde{H}$ is, in fact, entirely devoid of $a_R^\star({\bf k})$ and
$a_R({\bf k})$ operators except those that appear in the combination $\Gamma$,
so that the time-translation operator $\exp\left(-i\tilde{H}t\right)$ correctly
satisfies this requirement. Observable states in the alternative transformed
representation are described by state vectors in $\{|n\rangle\}$ which we
designate as $|N\rangle$. These observable states consist of massive photons,
electrons and positrons only, and have a positive norm. The operator
$\exp\left(-i\tilde{H}t\right)$ time translates such state vectors by
generating a new state vector, at a later time $t$, which consists of further
positive-norm state vectors $|N^\prime\rangle$, as well as additional ghost
states, all of which are represented by products of $a_Q^\star({\bf k})$
operators acting on positive-norm observable sets of states. At all times, the
positive-norm states alone just saturate unitarity. We can define a quotient
space consisting of the state vectors $|N\rangle$, which is the residue of
$\{|n\rangle\}$ after all zero-norm states have been excised from it. We can
also define another Hamiltonian $\tilde{H}_{\text{quot}}$, which consists of
those parts of $\tilde{H}$ that remain after we have removed all the terms in
which $a_Q^\star({\bf k})$ or $a_Q({\bf k})$ is a factor. This Hamiltonian is
given by
\begin{eqnarray}
\tilde{H}_{\text{quot}} &=& H_{e\bar{e}}+\sum_{\bf
k}\frac{\omega_k}{2}\left[a^\dagger({\bf k})a({\bf k}) + a({\bf
k})a^\dagger({\bf k})\right]\nonumber\\
&+&\sum_{\bf k}\frac{i\epsilon_{ln}k_l}{mk^2}\,j_n({\bf k})j_0(-{\bf k}) +
\sum_{\bf k}\frac{1}{2m^2}\,j_0({\bf k})j_0(-{\bf k})\nonumber\\
&+& \sum_{\bf k}\frac{3im^{3/2}k_l\phi({\bf k})}{16k^3}\,j_l({\bf k})j_0(-{\bf
k})\nonumber\\
&-&\sum_{\bf k}\frac{k}{m\sqrt{2\omega_k}}\left[a({\bf k})j_0(-{\bf k}) +
a^\dagger({\bf k})j_0({\bf k})\right]\nonumber\\
&+&\sum_{\bf k}\frac{\sqrt{\omega_k}k_l}{\sqrt{2}mk}\left[a({\bf k})j_l(-{\bf
k}) + a^\dagger({\bf k})j_l({\bf k})\right]\nonumber\\
&-&\sum_{\bf k}\frac{i\epsilon_{ln}k_n}{k\sqrt{2\omega_k}}\left[a({\bf
k})j_l(-{\bf k}) - a^\dagger({\bf k})j_l({\bf k})\right].
\end{eqnarray}
It is manifest that the state vectors $\exp\left(-i\tilde{H}t\right)|N\rangle$
and $\exp\left(-i\tilde{H}_{\text{quot}}t\right)|N\rangle$ have identical
projections on the set of state vectors $|N\rangle$ that define the quotient
space. The parts of $\tilde{H}$ that contain $a_Q^\star({\bf k})$ or $a_Q({\bf
k})$ as factors therefore do not play any role in the time evolution of state
vectors within the quotient space of observable states, and cannot have any
effect on the physical predictions of the theory.

To facilitate the comparison of the results we obtained in this covariant gauge
formulation of MCS theory with those derived in other gauges \cite{hl}, we make
another unitary transformation to a new representation so that operators
$\tilde{\cal P}$ are transformed to $\hat{\cal P}=e^{\Lambda}\tilde{\cal
P}e^{-\Lambda}$ where
\begin{equation}
\Lambda = \sum_{\bf k}\frac{k}{\sqrt{2}m\omega_k^{3/2}}\left[a({\bf
k})j_0(-{\bf k}) - a^\dagger({\bf k})j_0({\bf k})\right].
\label{eq:Lambda}
\end{equation}
The advantage of this representation lies in the fact that, after the
$\tilde{\cal P} \rightarrow \hat{\cal P}$ transformation, no interactions
remain in $\hat{H}$ that couple $j_0$ to $a({\bf k})$ or $a^\dagger({\bf k})$.
The ``static'' interactions of charged particles at rest with photons have
therefore been entirely absorbed into the nonlocal interactions among charge
and current densities. It is very convenient to normalize Hamiltonians in all
gauges to this common form, and to use the subspace $\{|n\rangle\}$ as the
Hilbert space in which the operators in the $\hat{\cal P}$ representation act.
In this new representation, the Hamiltonian is given by
\begin{eqnarray}
\hat{H} &=& H_0 + \sum_{\bf
k}\frac{im\epsilon_{ln}k_l}{k^2\omega_k^2}\,j_n({\bf k})j_0(-{\bf k}) +
\sum_{\bf k}\frac{1}{2\omega_k^2}\,j_0({\bf k})j_0(-{\bf k})\nonumber\\
&+& \sum_{\bf k}\frac{3im^{3/2}k_l\phi({\bf k})}{16k^3}\,j_l({\bf k})j_0(-{\bf
k})\nonumber\\
&+&\sum_{\bf k}\frac{mk_l}{\sqrt{2}k\omega_k^{3/2}}\left[a({\bf k})j_l(-{\bf
k}) + a^\dagger({\bf k})j_l({\bf k})\right]\nonumber\\
&-&\sum_{\bf k}\frac{i\epsilon_{ln}k_n}{k\sqrt{2\omega_k}}\left[a({\bf
k})j_l(-{\bf k}) - a^\dagger({\bf k})j_l({\bf k})\right]\nonumber\\
&-&\sum_{\bf k}\frac{4k^3}{m^{7/2}}\left[a_Q({\bf k})j_0(-{\bf k}) +
a_Q^\star({\bf k})j_0({\bf k})\right]\nonumber\\
&-&\sum_{\bf k}\frac{8ik\epsilon_{ln}k_n}{m^{5/2}}\left[a_Q({\bf k})j_l(-{\bf
k}) - a_Q^\star({\bf k})j_l({\bf k})\right]\nonumber\\
&+&\sum_{\bf k}\frac{4k^2k_l}{m^{7/2}}\left[a_Q({\bf k})j_l(-{\bf k}) +
a_Q^\star({\bf k})j_l({\bf k})\right]\nonumber\\
&+&(1-\gamma)\sum_{\bf k}\frac{2k}{m^{3/2}}\left[a_Q({\bf k})j_0(-{\bf k}) +
a_Q^\star({\bf k})j_0({\bf k})\right]\nonumber\\
&-&(1-\gamma)\sum_{\bf k}\frac{2k_l}{m^{3/2}}\left[a_Q({\bf k})j_l(-{\bf k}) +
a_Q^\star({\bf k})j_l({\bf k})\right]\nonumber\\
&+&\sum_{\bf k}k\theta({\bf k})\left[a_Q({\bf k})j_0(-{\bf k}) + a_Q^\star({\bf
k})j_0({\bf k})\right]\nonumber\\
&-&\sum_{\bf k}ik\phi({\bf k})\left[a_Q({\bf k})j_0(-{\bf k}) - a_Q^\star({\bf
k})j_0({\bf k})\right]\nonumber\\
&+&\sum_{\bf k}k_l\theta({\bf k})\left[a_Q({\bf k})j_l(-{\bf k}) +
a_Q^\star({\bf k})j_l({\bf k})\right]\nonumber\\
&+&\sum_{\bf k}ik_l\phi({\bf k})\left[a_Q({\bf k})j_l(-{\bf k}) -
a_Q^\star({\bf k})j_l({\bf k})\right].
\label{eq:hatH}
\end{eqnarray}
The similarly transformed fields are
\begin{equation}
\hat{A}_l({\bf x}) = A_l({\bf x}) - \sum_{\bf
k}\frac{im\epsilon_{ln}k_n}{k^2\omega_k^2}\,j_0({\bf k})e^{i{\bf k\cdot x}},
\end{equation}
\begin{equation}
\hat{\Pi}_l({\bf x}) = \Pi_l({\bf x}) + \sum_{\bf
k}\frac{ik_l(k^2+\omega_k^2)}{2k^2\omega_k^2}\,j_0({\bf k})e^{i{\bf k\cdot x}},
\end{equation}
\begin{eqnarray}
\hat{A}_0({\bf x}) &=& A_0({\bf x}) + (1-\gamma)\sum_{\bf k}
\frac{1}{4k^2}\,j_0({\bf k})e^{i{\bf k\cdot x}}\nonumber\\
&+& \sum_{\bf k}\frac{1}{2m^2}\,j_0({\bf k})e^{i{\bf k\cdot x}} - \sum_{\bf
k}\frac{m^{3/2}\theta({\bf k})}{8k^2}\,j_0({\bf k})e^{i{\bf k\cdot x}},
\end{eqnarray}
\begin{equation}
\hat{G}({\bf x}) = G({\bf x}) - \sum_{\bf k} \frac{i}{k}\,j_0({\bf k})e^{i{\bf
k\cdot x}},
\end{equation}
and the gauge-invariant $\hat{\psi}({\bf x})$
\begin{equation}
\hat{\psi}({\bf x}) = e^{{\cal D}^\prime({\bf x})}\psi({\bf x})
\end{equation}
where
\begin{equation}
{\cal D}^\prime({\bf x})=-ie\int d{\bf y}\ \left[\xi^\prime(|{\bf
x-y}|)\partial_lA_l({\bf x}) + \chi^\prime(|{\bf
x-y}|)\epsilon_{ln}\partial_lA_n({\bf x}) + \zeta^\prime(|{\bf x-y}|)G({\bf
x})\right]
\end{equation}
with
\begin{equation}
\xi^\prime(|{\bf x-y}|) = \sum_{\bf k}\frac{1}{2\omega_k^2}\,e^{i{\bf
k\cdot(x-y)}},
\end{equation}
\begin{equation}
\chi^\prime(|{\bf x-y}|) = \sum_{\bf k}\frac{1}{m\omega_k^2}\,e^{i{\bf
k\cdot(x-y)}},
\end{equation}
and
\begin{equation}
\zeta^\prime(|{\bf x-y}|) = \sum_{\bf k}\frac{k^2}{m^2\omega_k^2}\,e^{i{\bf
k\cdot(x-y)}}.
\end{equation}
The transformed electric and magnetic field are
\begin{equation}
\hat{E}_l({\bf x}) = E_l({\bf x}) + \hat{\cal E}_l({\bf x})
\end{equation}
and
\begin{equation}
\hat{B}({\bf x}) = B({\bf x}) + \hat{\cal B}({\bf x})
\end{equation}
where $E_l({\bf x})$ and $B({\bf x})$ are given by Eqs.~(\ref{eq:Elfield}) and
(\ref{eq:Bfield}), respectively, and
\begin{equation}
\hat{\cal E}_l({\bf x}) = -\frac{1}{2\pi}\,\frac{\partial}{\partial x_l}\,\int
d{\bf y}\ K_0(m|{\bf x-y}|)j_0({\bf y}),
\end{equation}
and
\begin{equation}
\hat{\cal B}({\bf x}) = -\frac{m}{2\pi}\,\int d{\bf y}\ K_0(m|{\bf
x-y}|)j_0({\bf y}).
\end{equation}

The quotient space Hamiltonian corresponding to the Hamiltonian $\hat{H}$ given
by Eq.~(\ref{eq:hatH}) can be written as
\begin{equation}
\hat{H}_{\text{quot}} = H_{e\bar{e}}+ \sum_{\bf
k}\frac{\omega_k}{2}\left[a^\dagger({\bf k})a({\bf k}) + a({\bf
k})a^\dagger({\bf k})\right] + \hat{H}_{\text{I}}
\label{eq:Hhat0}
\end{equation}
where
\begin{eqnarray}
\hat{H}_{\text{I}} &=& \int d{\bf x}\,d{\bf y}\ j_0({\bf
x})\epsilon_{ln}j_l({\bf y})(x-y)_n{\cal F}(|{\bf x-y}|)\nonumber\\
&+& \int d{\bf x}\,d{\bf y}\ j_0({\bf x})j_0({\bf y})K_0(m|{\bf
x-y}|)\nonumber\\
&+& \sum_{\bf k}\frac{3im^{3/2}k_l\phi({\bf k})}{16k^3}\,j_l({\bf k})j_0(-{\bf
k})\nonumber\\
&+&\sum_{\bf k}\frac{mk_l}{\sqrt{2}k\omega_k^{3/2}}\left[a({\bf k})j_l(-{\bf
k}) + a^\dagger({\bf k})j_l({\bf k})\right]\nonumber\\
&-&\sum_{\bf k}\frac{i\epsilon_{ln}k_n}{k\sqrt{2\omega_k}}\left[a({\bf
k})j_l(-{\bf k}) - a^\dagger({\bf k})j_l({\bf k})\right]
\label{eq:hatHI}
\end{eqnarray}
and where $K_0(x)$ is a modified Bessel function and\footnote{The definition of
${\cal F}(R)$ given in Ref.~\cite{hl} is lacking a factor ${1}/{2\pi}$.}
\begin{equation}
{\cal F}(R) = -\frac{m}{2\pi}\int_0^\infty du\ \frac{J_1(u)}{u^2+(mR)^2}.
\end{equation}
We observe that ${\cal F}(R)$ approaches the limits
\begin{equation}
\lim_{mR\rightarrow 0} {\cal F}(R) = \frac{1}{4\pi R}
\end{equation}
and
\begin{equation}
{\cal F}(R) \rightarrow \frac{1}{2\pi mR^2}
\label{eq:calf(r)}
\end{equation}
as $mR\rightarrow\infty$. The interaction Hamiltonian $\hat{H}_{\text{I}}$
describes the interaction of massive photons with charged fermions. It also
describes nonlocal interactions between charged fermions. These interactions
include the $(2+1)$-dimensional analogue of the Coulomb interaction, with the
inverse power of distance between charges replaced by the modified Bessel
function $K_0(m|{\bf x-y}|)$. Another such interaction, which has no analogue
in QED$_4$, couples charges and the transverse components of currents. The
expressions for the nonlocal interactions among charge and current densities
that result from the elimination of ``ghost'' components of the gauge fields,
are well behaved and free from the kind of infrared singularities that one
might anticipate from massless particle exchange in a $(2+1)$-dimensional
model. The Hamiltonian $\hat{H}_{\text{I}}$, which is obtained from the
implementation of Gauss's law and the gauge condition $\partial_\mu A^\mu =0$,
is identical to the Hamiltonian we obtained previously by implementing Gauss's
law and the gauge condition $A_0 = 0$. This identity extends also to the
electric field, $\hat{E}_l({\bf x})$, and the magnetic field, $\hat{B}({\bf
x})$, which are identical to the corresponding expressions for the electric and
magnetic fields in the $A_0=0$ gauge. These identities make the gauge
invariance of this theory very manifest, because we have eliminated physically
meaningless differences in form that arise when unitary equivalence between
sets of dynamical variables have not been fully recognized and used to
demonstrate gauge equivalence. Later in this paper we will extend this gauge
equivalence to the Coulomb gauge.

The term
\begin{equation}
h = \sum_{\bf k}\frac{3im^{3/2}k_l\phi({\bf k})}{16k^3}\,j_l({\bf k})j_0(-{\bf
k})
\end{equation}
that appears in the Hamiltonian given by Eq.~(\ref{eq:hatH}) is a total
time-derivative which can be expressed as $h = i[H_0,\chi]$ or as $h =
i[\tilde{H},\chi]$ where
\begin{equation}
\chi = -\sum_{\bf k}\frac{3m^{3/2}\phi({\bf k})}{32k^3}\,j_0({\bf k})j_0(-{\bf
k}).
\end{equation}
The fact that $h$ is a total time-derivative gives us {\em a priori} confidence
that it will not affect the $S$-matrix produced by this theory. A formal
argument that confirms this result has been given previously and will not be
repeated here \cite{khqedtemp}.

The fact that $h$ is a total time-derivative of $\chi$, and that $\chi$ and $h$
commute, establishes the relationship
\begin{equation}
e^{-ie\chi}\hat{H}e^{ie\chi}-\hat{H} = e^{-ie\chi}\tilde{H}e^{ie\chi}-\tilde{H}
= e^{-ie\chi}H_0e^{ie\chi}-H_0 = h.
\end{equation}
This, in turn, demonstrates that if we combine the Hamiltonian $\hat{H}$ with
the Fock space $\{|n\rangle\}$, the resulting formalism will be unitarily
equivalent to the Hamiltonian $\hat{H}+h$ combined with the Hilbert space
$\{e^{-ie\chi}|n\rangle\}$. The choice of a Hilbert space in which a
Hamiltonian and the other dynamical variables of a model are to act, is an
independent assumption in the axiomatic structure of the theory. We could
equally well have chosen to combine the Hamiltonian $H+h$ with the Fock space
$\{|n\rangle\}$, simply by choosing a nonvanishing $\phi({\bf k})$. The
$S$-matrix would not have been affected by that substitution, but there would
have been changes in the time-evolution of state vectors at times $t$ remote
from the asymptotic regions $t\rightarrow\pm\infty$; the effects of these
changes in time-evolution cancel by the time the asymptotic region
$t\rightarrow\infty$ is reached. It would be desirable to have a ``natural''
principle for associating Hamiltonians and Hilbert spaces, but when the
substitution of one Hamiltonian for another has no effect on the $S$-matrix,
there are no physical reasons for preferring one combination over the other. We
have used a ``minimal'' principle in our work, somewhat in the spirit of the
``minimal coupling'' rule for coupling gauge fields to matter. This minimal
principle dictates that parts of Hamiltonians, like $h$, that make no
contribution at all to the $S$-matrix, are excluded in representations in which
the Fock space $\{|n\rangle\}$ represents the states that implement Gauss's law
and the gauge condition. This principle does not help to make a selection in
every case, but it answers the need adequately in the case of Abelian gauge
theories.

\section{The perturbative regime}
The perturbative theory involves the vertices dictated by the interaction
Hamiltonian given in Eq.~(\ref{eq:Hinteraction}) and the propagators for the
interaction-picture operators $\psi(x)$, $\bar{\psi}(x)$, and $A^\mu(x)$
obtained from ${\cal P}(x) = \exp\left(iH_0t\right){\cal P}({\bf
x})\exp\left(-iH_0t\right)$. The gauge fields in the interaction picture are
found to be
\begin{eqnarray}
A_l(x) &=& -\sum_{\bf k}\frac{\sqrt{\omega_k}k_l}{\sqrt{2}mk}\left[a({\bf
k})e^{-ik_\mu x^\mu}+a^\dagger({\bf k})e^{ik_\mu x^\mu}\right]\nonumber\\
&+&\sum_{\bf k}\frac{i\epsilon_{ln}k_n}{k\sqrt{2\omega_k}}\left[a({\bf
k})e^{-ik_\mu x^\mu}-a^\dagger({\bf k})e^{ik_\mu x^\mu}\right]\nonumber\\
&+&\sum_{\bf k}\frac{8ik\epsilon_{ln}k_n}{m^{5/2}}\left[a_Q({\bf
k})e^{-ik^\prime_\mu x^\mu}-a_Q^\star({\bf k})e^{ik^\prime_\mu
x^\mu}\right]\nonumber\\
&+& (1-\gamma)\sum_{\bf k}\frac{2k_l}{m^{3/2}}\left[a_Q({\bf
k})e^{-ik^\prime_\mu x^\mu}+a_Q^\star({\bf k})e^{ik^\prime_\mu
x^\mu}\right]\nonumber\\
&-&\sum_{\bf k}\frac{4k^2k_l}{m^{7/2}}\left[a_Q({\bf k})e^{-ik^\prime_\mu
x^\mu}+a_Q^\star({\bf k})e^{ik^\prime_\mu x^\mu}\right]\nonumber\\
&+&\sum_{\bf k}\frac{m^{3/2}k_l}{16k^3}\left[a_R({\bf k})e^{-ik^\prime_\mu
x^\mu}+a_R^\star({\bf k})e^{ik^\prime_\mu x^\mu}\right]\nonumber\\
&+&(1-\gamma)\sum_{\bf k}\frac{4ix_0kk_l}{m^{3/2}}\left[a_Q({\bf
k})e^{-ik^\prime_\mu x^\mu}-a_Q^\star({\bf k})e^{ik^\prime_\mu x^\mu}\right]
\end{eqnarray}
and
\begin{eqnarray}
A_0(x) &=& -\sum_{\bf k}\frac{k}{m\sqrt{2\omega_k}}\left[a({\bf k})e^{-ik_\mu
x^\mu}+a^\dagger({\bf k})e^{ik_\mu x^\mu}\right]\nonumber\\
&-&\sum_{\bf k}\frac{4k^3}{m^{7/2}}\left[a_Q({\bf k})e^{-ik^\prime_\mu
x^\mu}+a_Q^\star({\bf k})e^{ik^\prime_\mu x^\mu}\right]\nonumber\\
&-&(1-\gamma)\sum_{\bf k}\frac{2k}{m^{3/2}}\left[a_Q({\bf k})e^{-ik^\prime_\mu
x^\mu}+a_Q^\star({\bf k})e^{ik^\prime_\mu x^\mu}\right]\nonumber\\
&+&\sum_{\bf k}\frac{m^{3/2}}{16k^2}\left[a_R({\bf k})e^{-ik^\prime_\mu
x^\mu}+a_R^\star({\bf k})e^{ik^\prime_\mu x^\mu}\right]\nonumber\\
&+&(1-\gamma)\sum_{\bf k}\frac{4ix_0k^2}{m^{3/2}}\left[a_Q({\bf
k})e^{-ik^\prime_\mu x^\mu}-a_Q^\star({\bf k})e^{ik^\prime_\mu x^\mu}\right]
\end{eqnarray}
where $k_\mu x^\mu = \omega_kx_0 - {\bf k\cdot x}$ and $k_\mu^\prime x^\mu =
kx_0 - {\bf k\cdot x}$. We use $A_l(x)$ and $A_0(x)$ in the expression for the
propagator,
\begin{equation}
D^{\mu\nu}(x,y)=\langle 0|{\sf T}(A^\mu(x)A^\nu(y))|0\rangle
\end{equation}
where ${\sf T}$ designates time-ordering, and where $|0\rangle$ is the
perturbative vacuum annihilated by all annihilation operators, $a({\bf k})$,
$a_Q({\bf k})$ and $a_R({\bf k})$, as well as, $e({\bf k})$ and $\bar{e}({\bf
k})$ for electrons and positrons, respectively. We obtain the expressions
\begin{eqnarray}
D_{ln}(x,y) &=& (1-\gamma)\sum_{\bf
k}\frac{k_lk_n}{4k^3}\left[e^{-ik^\prime_\mu(x-y)^\mu}\Theta(x_0-y_0) +
e^{ik^\prime_\mu(x-y)^\mu}\Theta(y_0-x_0)\right]\nonumber\\
&+&(1-\gamma)\sum_{\bf
k}\frac{ik_lk_n(x_0-y_0)}{4k^2}\left[e^{-ik^\prime_\mu(x-y)^\mu}\Theta(x_0-y_0)
- e^{ik^\prime_\mu(x-y)^\mu}\Theta(y_0-x_0)\right]\nonumber\\
&-&\sum_{\bf
k}\frac{k_lk_n}{2m^2k}\left[e^{-ik^\prime_\mu(x-y)^\mu}\Theta(x_0-y_0) +
e^{ik^\prime_\mu(x-y)^\mu}\Theta(y_0-x_0)\right]\nonumber\\
&+&\sum_{\bf
k}\frac{i\epsilon_{ln}}{2m}\left[e^{-ik^\prime_\mu(x-y)^\mu}\Theta(x_0-y_0) -
e^{ik^\prime_\mu(x-y)^\mu}\Theta(y_0-x_0)\right]\nonumber\\
&-&\sum_{\bf
k}\frac{i\epsilon_{ln}}{2m}\left[e^{-ik_\mu(x-y)^\mu}\Theta(x_0-y_0) -
e^{ik_\mu(x-y)^\mu}\Theta(y_0-x_0)\right]\nonumber\\
&+&\sum_{\bf k}\frac{1}{2\omega_k}\left[\delta_{ln}+
\frac{k_lk_n}{m^2}\right]
\left[e^{-ik_\mu(x-y)^\mu}\Theta(x_0-y_0) -
e^{ik_\mu(x-y)^\mu}\Theta(y_0-x_0)\right],
\end{eqnarray}
\begin{eqnarray}
D_{0l}(x,y) &=& (1-\gamma)\sum_{\bf
k}\frac{ik_l(x_0-y_0)}{4k}\left[e^{-ik^\prime_\mu(x-y)^\mu}\Theta(x_0-y_0) -
e^{ik^\prime_\mu(x-y)^\mu}\Theta(y_0-x_0)\right]\nonumber\\
&-&\sum_{\bf k}\frac{k_l}{2m^2}\left[e^{-ik^\prime_\mu(x-y)^\mu}\Theta(x_0-y_0)
+ e^{ik^\prime_\mu(x-y)^\mu}\Theta(y_0-x_0)\right]\nonumber\\
&-&\sum_{\bf
k}\frac{i\epsilon_{ln}k_n}{2mk}\left[e^{-ik^\prime_\mu(x-y)^\mu}\Theta(x_0-y_0)
- e^{ik^\prime_\mu(x-y)^\mu}\Theta(y_0-x_0)\right]\nonumber\\
&+&\sum_{\bf k}\frac{k_l}{2m^2}\left[e^{-ik_\mu(x-y)^\mu}\Theta(x_0-y_0)+
e^{ik_\mu(x-y)^\mu}\Theta(y_0-x_0)\right]\nonumber\\
&+&\sum_{\bf
k}\frac{i\epsilon_{ln}k_n}{2m\omega_k}\left[e^{-ik_\mu(x-y)^\mu}\Theta(x_0-y_0)
- e^{ik_\mu(x-y)^\mu}\Theta(y_0-x_0)\right],
\end{eqnarray}
and
\begin{eqnarray}
D_{00}(x,y) &=& -(1-\gamma)\sum_{\bf
k}\frac{1}{4k}\left[e^{-ik^\prime_\mu(x-y)^\mu}\Theta(x_0-y_0) +
e^{ik^\prime_\mu(x-y)^\mu}\Theta(y_0-x_0)\right]\nonumber\\
&+&(1-\gamma)\sum_{\bf
k}\frac{i(x_0-y_0)}{4}\left[e^{-ik^\prime_\mu(x-y)^\mu}\Theta(x_0-y_0) -
e^{ik^\prime_\mu(x-y)^\mu}\Theta(y_0-x_0)\right]\nonumber\\
&-&\sum_{\bf k}\frac{k}{2m^2}\left[e^{-ik^\prime_\mu(x-y)^\mu}\Theta(x_0-y_0) +
e^{ik^\prime_\mu(x-y)^\mu}\Theta(y_0-x_0)\right]\nonumber\\
&+&\sum_{\bf
k}\frac{k^2}{2m^2\omega_k}\left[e^{-ik_\mu(x-y)^\mu}\Theta(x_0-y_0) +
e^{ik_\mu(x-y)^\mu}\Theta(y_0-x_0)\right]
\end{eqnarray}
which can be represented as
\begin{equation}
D^{\mu\nu}(x,y) = -i\int \frac{d^3k}{(2\pi)^3}\
D^{\mu\nu}(k)e^{-ik_\mu(x-y)^\mu}
\label{eq:Dmunuxy}
\end{equation}
where
\begin{eqnarray}
D^{\mu\nu}(k) &=& (1-\gamma)\,\frac{k^\mu k^\nu}{(k_\alpha k^\alpha +
i\epsilon)^2} - \frac{k^\mu k^\nu}{(k_\alpha k^\alpha + i\epsilon)(k_\beta
k^\beta - m^2 + i\epsilon)}\nonumber\\
&+& \frac{g^{\mu\nu}}{k_\alpha k^\alpha - m^2 + i\epsilon} +
\frac{im\epsilon^{\mu\nu\lambda}k_\lambda}{(k_\alpha k^\alpha +
i\epsilon)(k_\beta k^\beta - m^2 + i\epsilon)};
\label{eq:DmunuMCS}
\end{eqnarray}
in Eqs.~(\ref{eq:Dmunuxy}) and (\ref{eq:DmunuMCS}), $k_0$, $k_1$ and $k_2$ are
independent variables.

\section{Poincar\'e structure}
\label{sec:poincare}
The consistency of the canonical formulation of this model can be given further
support by constructing the canonical Poincar\'e generators, and using the
canonical commutation rules given in Sec.~\ref{sec:formulation} to demonstrate
that they implement the required algebra. In $2+1$ dimensions, the Poincar\'e
group has six generators: one $(J)$ for rotation, two $(K_l)$ for boosts, and
two $(P_l)$ and one $(P_0)$ for space and time translations, respectively. The
translation operators $P_l$ and $P_0$ can be written as
\begin{equation}
P_l = \int d{\bf x}\ {\cal P}_l({\bf x})
\label{eq:Pl}
\end{equation}
and
\begin{equation}
P_0 = \int d{\bf x}\ {\cal P}_0({\bf x})
\label{eq:P0}
\end{equation}
where ${\cal P}_0({\bf x}) = {\cal H}({\bf x})$, with ${\cal H}({\bf x})$ given
by Eq.~(\ref{eq:calH}), and the canonical form of ${\cal P}_l$ is given by
\begin{equation}
{\cal P}_l({\bf x}) = G({\bf x})\partial_lA_0({\bf x}) - \Pi_n({\bf
x})\partial_lA_n({\bf x})-i\psi^\dagger({\bf x})\partial_l\psi({\bf x}).
\end{equation}
Similarly, we follow the canonical procedure \cite{bjorken} and express the
rotation and boost operators $J$ and $K_l$, respectively, as
\begin{equation}
J = \int d{\bf x}\ \epsilon_{ln}x_l{\cal P}_n({\bf x}) + \int d{\bf x}\
\kappa_{\text{rotation}}({\bf x})
\label{eq:J}
\end{equation}
and
\begin{equation}
K_l = x_0P_l - \int d{\bf x}\ x_l{\cal P}_0({\bf x}) + \int d{\bf x}\
\kappa^{\text{boost}}_l({\bf x})
\label{eq:Kl}
\end{equation}
where
\begin{equation}
\kappa_{\text{rotation}} = \epsilon_{ln}A_l\Pi_n - \case 1/2
\psi^\dagger\gamma_0\psi
\end{equation}
and
\begin{equation}
\kappa^{\text{boost}}_l = -A_lG + A_0\Pi_l + \case 1/2
i\psi^\dagger\gamma_0\gamma_l\psi.
\end{equation}
The term $\kappa_{\text{rotation}}$ implements the mixing of the space
components of the fields during a rotation. It arises from the fact that, under
an infinitesimal rotation $\delta\theta$ about an axis perpendicular to the 2-D
plane, the components of $A^\mu$ change as follows:
\begin{equation}
\delta A_l({\bf x}) = -[\epsilon_{ij}x_i\partial_jA_l({\bf x}) +
\epsilon_{ln}A_n({\bf x})]\,\delta\theta
\label{eq:deltaAl}
\end{equation}
and
\begin{equation}
\delta A_0({\bf x}) = -\epsilon_{ij}x_i\partial_jA_0({\bf x})\,\delta\theta;
\end{equation}
the spinor field transforms under rotations as
\begin{equation}
\delta\psi({\bf x}) = -[\epsilon_{ij}x_i\partial_j\psi({\bf x}) - \case 1/2
i\gamma_0\psi({\bf x})]\,\delta\theta,
\end{equation}
so that $\delta(\bar{\psi}\gamma_l\psi) =
-[\epsilon_{ij}x_i\partial_j(\bar{\psi}\gamma_l\psi) +
\epsilon_{ln}(\bar{\psi}\gamma_n\psi)]\,\delta\theta$. Similarly, the term
$\kappa^{\text{boost}}_l$ in $K_l$ mixes the space-time components of the
fields under a boost. For example, under an infinitesimal boost $\delta\beta_l$
along the $l$-direction, the components of $A^\mu$ transform as follows
\begin{equation}
\delta A_0({\bf x}) = -[x_0\partial_lA_0({\bf x}) + x_l\partial_0A_0({\bf x})-
A_l({\bf x})]\,\delta\beta_l
\label{eq:deltaA0betal}
\end{equation}
and
\begin{equation}
\delta A_i({\bf x}) = -[x_0\partial_lA_i({\bf x}) + x_l\partial_0A_i({\bf x})-
\delta_{il}A_0({\bf x})]\,\delta\beta_l.
\label{eq:deltaAibetal}
\end{equation}
The spinor field transforms as
\begin{equation}
\delta\psi({\bf x}) = -[x_0\partial_l\psi({\bf x}) + x_l\partial_0\psi({\bf x})
- \case 1/2 i\gamma_0\gamma_l\psi({\bf x})]\,\delta\beta_l.
\label{eq:deltapsi}
\end{equation}
Using Eq.~(\ref{eq:Kl}), we can also demonstrate the mixing of the electric
field $E_l$ and the magnetic field $B$ under a Lorentz transformation:
\begin{equation}
\delta E_l = -[x_0\partial_nE_l + x_n\partial_0E_l
+\epsilon_{ln}B]\,\delta\beta_n
\end{equation}
and
\begin{equation}
\delta B = -[x_0\partial_nB + x_n\partial_0B +
\epsilon_{ln}E_l]\,\delta\beta_n.
\end{equation}

Use of the canonical commutation rules leads to the following commutation rules
for the Poincar\'e generators:
\begin{equation}
[P_l,P_n]=0,
\label{eq:PlPn}
\end{equation}
\begin{equation}
[H,P_l] = [H,J]=0,
\label{eq:PlJH}
\end{equation}
\begin{equation}
[H,K_l]=iP_l,
\label{eq:HKl}
\end{equation}
\begin{equation}
[P_l,K_n] = i\delta_{ln}H,
\label{eq:PlKn}
\end{equation}
\begin{equation}
[P_l,J] = -i\epsilon_{ln}P_n,
\label{eq:PlJ}
\end{equation}
\begin{equation}
[J,K_l] = i\epsilon_{ln}K_n,
\label{eq:JKl}
\end{equation}
and
\begin{equation}
[K_l,K_n] = -i\epsilon_{ln}J.
\label{eq:KlKn}
\end{equation}
We observe that these commutation rules form a closed Lie algebra, and that
they are consistent with the transformations given in
Eqs.~(\ref{eq:deltaAl})--(\ref{eq:deltapsi}).\footnote{The presence of $x_l$ in
$\int d{\bf x}\ x_l{\cal P}_0({\bf x})$ introduces ambiguities into the
expression for the commutator of two boost operators, that appear when the
momentum space representations of the gauge fields are used. We will discuss
these ambiguities in an appendix to this paper.} The angular momentum, which is
an axial vector in three-dimensional space, degenerates into a scalar in two
dimensions. All spatial and temporal displacements commute; momentum and
angular momentum are time-displacement invariant. Equations~(\ref{eq:HKl}) and
(\ref{eq:PlKn}) express the infinitesimal Lorentz transformation in $2+1$
dimensions.

The consistency of the Lie algebra formed by these canonical Poincar\'e
generators for this model supports the use we make of the angular momentum,
$J$, to implement rotations. We will discuss this topic in the next section.

\section{Anomalous rotation and exotic statistics}
\label{sec:anomalousrotation}
It has been demonstrated that in pure CS theory a charged particle of charge
$e$, interacting with a CS field in the absence of the Maxwell kinetic energy,
can acquire the phase $e^2/m$ when it is rotated through $2\pi$ radians
\cite{swanson}. The occurrence of the arbitrary phase has been attributed by
some authors to the imposition of Gauss's law in the CS theory \cite{CSgauss}.
In Ref.~\cite{anyon}, we have shown that the rotational anomaly that arises in
CS theory has nothing to do with the implementation of Gauss's law. We have
constructed charged fermion states in MCS theory in the temporal gauge which
rotate normally---i.e., they change sign under a $2\pi$ rotation---even when
these charged states implement Gauss's law. We have also demonstrated that
these charged states in MCS theory obey standard Fermi-Dirac statistics
although they obey Gauss's law \cite{hl}. By pursing the same analysis detailed
in Ref.~\cite{hl}, we will confirm that conclusion in the formulation of MCS
theory in covariant gauges. Furthermore, we will show that other states can
also be constructed that satisfy Gauss's law, and that do acquire an arbitrary
phase under $2\pi$ rotations. These states also obey standard Fermi-Dirac
statistics and give rise to the same $S$-matrix elements as the states that
rotate normally.

The rotation operator we will use is $R({\theta})=e^{iJ\theta}$ where $J$ is
the canonical (Noether) angular momentum given by Eq.~(\ref{eq:J}), which we
have just identified as one of the six Poincar\'e generators that close under
the commutator algebra required for the Poincar\'e group. We will express $J$
as
\begin{equation}
J = J_{\text{g}} + J_{\text{e}}
\end{equation}
with
\begin{equation}
 J_{\text{g}} = -\int d{\bf x}\ \Pi_lx_i\epsilon_{ij}\partial_jA_l + \int d{\bf
x}\ Gx_i\epsilon_{ij}\partial_jA_0 - \int d{\bf x}\ \epsilon_{ij}\Pi_iA_j
\label{eq:Jg}
\end{equation}
and
\begin{equation}
J_{\text{e}} = -i\int d{\bf x}\ \psi^\dagger x_i\epsilon_{ij}\partial_j\psi -
\case 1/2 \int d{\bf x}\ \psi^\dagger\gamma_0\psi.
\label{eq:Je}
\end{equation}
As pointed out in the previous section, $J$ is time independent since
$[H,J]=0$.

The interpretation of these angular momentum operators in terms of the angular
momenta of the constituent particle-mode excitations is greatly simplified when
single-particle plane waves are replaced with eigenstates of angular momentum.
We therefore substitute gauge-field annihilation and creation operators
describing excitations with definite angular momentum, $\alpha_n(k)$ and
$\alpha_n^\dagger(k)$, respectively, for the corresponding plane-wave
excitations $a({\bf k})$ and $a^\dagger({\bf k})$. This is accomplished by
using
\begin{equation}
\alpha_n(k) = \frac{e^{in\pi/2}}{2\pi}\int d{\tau}\ a({\bf k})e^{-in\tau}
\end{equation}
where $\tau$ is the angle that fixes the direction of ${\bf k}$ in the plane; a
corresponding expression relates the Hermitian adjoints $\alpha_n^\dagger(k)$
and $a^\dagger({\bf k})$ \cite{hl}. Similarly, we can define the following
single-particle solutions of the Dirac equation in polar coordinates,
\begin{equation}
u_+(n,k;\rho,\phi) = \frac{1}{\sqrt{2\bar{\omega}_k(M+\bar{\omega}_k)}}\,
\left[\begin{array}{c}ikJ_n(k\rho)e^{in\phi} \\
(M+\bar{\omega}_k)J_{n+1}(k\rho)e^{i(n+1)\phi}\end{array}\right]
\label{eq:u+}
\end{equation}
and
\begin{equation}
u_-(n,k;\rho,\phi) = \frac{1}{\sqrt{2\bar{\omega}_k(M+\bar{\omega}_k)}}\,
\left[\begin{array}{c}(M+\bar{\omega}_k)J_n(k\rho)e^{in\phi}\\
ikJ_{n+1}(k\rho)e^{i(n+1)\phi}\end{array}\right],
\label{eq:u-}
\end{equation}
where $J_s(x)$ is the Bessel function of order $s$. Using $u_\pm$, we can
represent $\psi$ in the angular momentum representation as
\begin{equation}
\psi(\rho,\phi) = \sum_{n,k}[b_n(k)u_+(n,k;\rho,\phi) +
\bar{b}_n^\dagger(k)u_-(n,k;\rho,\phi)],
\end{equation}
where $\sum_{n,k}=\sum_n\int {k\,dk}/{2\pi}$; $b_n(k)$ and $\bar{b}_n(k)$ are
the electron and positron annihilation operators, respectively, for states with
definite angular momentum; $b_n^\dagger(k)$ and $\bar{b}_n^\dagger(k)$ are the
corresponding creation operators. The operator $b_n({\bf k})$ is related to
$e({\bf k})$ by
\begin{equation}
b_n(k) = \frac{e^{in\pi/2}}{2\pi}\int d{\tau}\ e({\bf k})e^{-in\tau}.
\end{equation}
Similar expressions relate $\bar{b}_n({\bf k})$ to $\bar{e}({\bf k})$, the
adjoints $b_n^\dagger({\bf k})$ to $e^\dagger({\bf k})$, and
$\bar{b}_n^\dagger({\bf k})$ to $\bar{e}^\dagger({\bf k})$. To write the
angular momentum operator $J$ in terms of annihilation and creation operators
of states with definite angular momentum, we first write the gauge fields
$A_l$, $\Pi_l$, $A_0$, $G$ and the spinor fields $\psi$ and $\psi^\dagger$ in
terms of $\alpha_n(k)$, $\alpha_{Q,n}(k)$, $\alpha_{R,n}(k)$, $b_n(k)$,
$\bar{b}_n(k)$ and their adjoints, and then make the appropriate substitutions
in Eqs.~(\ref{eq:Jg}) and (\ref{eq:Je}). The resulting expressions for the
angular momenta $J_{\text{g}}$ and $J_{\text{e}}$ are
\begin{equation}
J_{\text{g}} = \sum_{n,k} n\,[\alpha_n^\dagger(k)\alpha_n(k) +
\alpha_{R,n}^\star\alpha_{Q,n}(k) + \alpha_{Q,n}^\star\alpha_{R,n}(k)]
\label{eq:Jgangular}
\end{equation}
and
\begin{equation}
J_{\text{e}} = \sum_{n,k} (n + \case 1/2)\,[b^\dagger_n(k)b_n(k) -
\bar{b}_n^\dagger(k)\bar{b}_n(k)].
\label{eq:Jeangular}
\end{equation}
Thus, the eigenvalues of $J$ are integral for a photon state, and half-integral
for an electron or positron state. We can also show, using
Eqs.~(\ref{eq:Jgangular}) and (\ref{eq:Jeangular}), that the rotation operator
$R(\theta) = e^{iJ\theta}$ rotates particle states correctly, e.g., the
electron state $|N\rangle = b_N^\dagger|0\rangle$ is rotated into
$|N^\prime\rangle = \exp\,[i(N+\case 1/2)\theta]|N\rangle$.

To investigate the rotational properties of charged states that obey the
Gauss's law constraint under $2\pi$ rotations, let us consider the ``bare''
one-electron state $|N\rangle=e^\dagger({\bf k})|0\rangle$. The one-electron
state $|N\rangle$ does not satisfy the constraint $\Omega({\bf k})|N\rangle
=0$, and thus is not in the physical subspace $\{|\nu\rangle\}$ defined in
Sec.~\ref{sec:implofgausslaw}. The electron state which satisfies
Eq.~(\ref{eq:omegaspace}) is given by $e^{D}|N\rangle$. It is therefore the
rotation of the state $e^{D}|N\rangle$ that we must analyze and not $|N\rangle$
when we use the untransformed representation of the states, fields and
dynamical variables; in the untransformed representation, the subspace
$\{|n\rangle\}$ represents states for which Gauss's law still remains to be
implemented. The rotation of the state $e^{D}|N\rangle$, which obey Gauss's law
in this untransformed representation, can be evaluated by writing
$R({\theta})e^{D}|N\rangle = e^{D}e^{-D}R({\theta})e^{D}|N\rangle =
e^{D}\tilde{R}(\theta)|N\rangle$. The rotation operator $\tilde{R}(\theta)$ is
given by $\tilde{R}(\theta)=e^{i\tilde{J}\theta}$ where $\tilde{J}$ is given by
\begin{equation}
\tilde{J} = J + {\cal J}
\end{equation}
where
\begin{eqnarray}
{\cal J} &=& -\sum_{\bf
k}\frac{m^{3/2}}{16k^3}\,\epsilon_{ln}k_l\frac{\partial\phi({\bf k})}{\partial
k_n}\,j_0(-{\bf k})j_0({\bf k})\nonumber\\
&-& \sum_{\bf k} i\epsilon_{ln}k_l\frac{\partial\theta({\bf k})}{\partial
k_n}\left[a_Q({\bf k})j_0(-{\bf k}) - a_Q^\star({\bf k})j_0({\bf
k})\right]\nonumber\\
&+& \sum_{\bf k}\epsilon_{ln}k_l\frac{\partial\phi({\bf k})}{\partial
k_n}\left[a_Q({\bf k})j_0(-{\bf k}) + a_Q^\star({\bf k})j_0({\bf k})\right].
\end{eqnarray}
The rotation operator $\tilde{R}(\theta)$ can also be obtained by noting that
in the alternate, transformed representation, $|N\rangle$ represents the
one-electron state that does implement Gauss's law and the gauge condition. In
that representation, all dynamical variables are represented by the
correspondingly transformed operators, so that the rotation operator is given
by $\tilde{R}=e^{-D}Re^{D}$. Equation~(\ref{eq:DQ}) reminds us that the forms
of $\phi({\bf k})$ and $\theta({\bf k})$ may be chosen arbitrarily without
disturbing the implementation of Gauss's law. In particular, $\phi({\bf k})$
and/or $\theta({\bf k})$ may be set to zero. In that case, $\tilde{J}=J$, and
the states that implement Gauss's law will rotate like the ``bare'' fermion
states that don't obey Gauss's law, i.e., they will change sign  in a $2\pi$
rotation. Other choices for $\phi({\bf k})$ and/or $\theta({\bf k})$ will lead
to different rotational properties for the charged states that obey Gauss's
law. If we choose
\begin{equation}
\phi({\bf k}) =
-\frac{8k^3}{m^{5/2}}\,\frac{\delta(k)}{k}\,\tan^{-1}\frac{k_2}{k_1},
\label{eq:phiforcalJ}
\end{equation}
and if we assume that we can carry out the integration over $d{\bf k}$ while
$j_0({\bf k})$ is still operator-valued, then the first term in ${\cal J}$
becomes $Q^2/4\pi m$, where $Q$ is the electron charge
\begin{equation}
Q = e\sum_{n,k}[b_n^\dagger(k)b_n(k)-\bar{b}^\dagger_n(k)\bar{b}_n(k)].
\end{equation}
Hence, under a $2\pi$ rotation, the state $e^D|N\rangle$ which obeys Gauss's
law picks up an arbitrary phase $e^{e^2/4\pi m}$, that is,
\begin{equation}
R(2\pi)e^{D}|N\rangle = -e^{e^2/4\pi m}e^D|N\rangle.
\end{equation}

Another important question to examine is whether ``exotic'' fractional
statistics develop when Gauss's law is implemented for charged states. If the
anticommutators for the spinor fields that implement Gauss's law,
$\{\tilde{\psi}({\bf x}),\tilde{\psi}^\dagger({\bf y})\}$ and
$\{\tilde{\psi}({\bf x}),\tilde{\psi}({\bf y})\}$, differ from the canonical
spinor anticommutators $\{\psi({\bf x}),\psi^\dagger({\bf y})\}=\delta({\bf
x-y})$ and $\{\psi({\bf x}),\psi({\bf y})\}=0$, then that difference may signal
that the excitations of $\tilde{\psi}$ and $\tilde{\psi}^\dagger$ are subject
to fractional statistics.  To show that the anticommutation rules for
$\tilde{\psi}$ and $\tilde{\psi}^\dagger$ are identical to the anticommutation
rules for the unconstrained $\psi$ and $\psi^\dagger$, we refer to
Eq.~(\ref{eq:tildepsi}). Since the gauge and the spinor fields commute at equal
times, then ${\cal D}({\bf x})$ commutes with $\psi({\bf x})$ and therefore
\begin{equation}
\{\tilde{\psi}({\bf x}),\tilde{\psi}^\dagger({\bf y})\} = \delta({\bf x-y})
\label{eq:X1}
\end{equation}
and
\begin{equation}
\{\tilde{\psi}({\bf x}),\tilde{\psi}({\bf y})\} = 0.
\label{eq:X2}
\end{equation}
The gauge-independent fields $\tilde{\psi}$ and $\tilde{\psi}^\dagger$ obey the
same anticommutation rules as the gauge-dependent $\psi$ and $\psi^\dagger$,
and are not subject to any exotic graded anticommutator algebra. The same
conclusion also follows from the observation that $\tilde{\psi}({\bf x})$ and
$\tilde{\psi}^\dagger({\bf x})$ are unitary transforms of $\psi({\bf x})$ and
$\psi^\dagger({\bf x})$ respectively, and that the anticommutator algebra for
fermion fields is invariant to the unitary transformation. The electron and
positron states that implement Gauss's law therefore obey standard Fermi---not
fractional---statistics. For example, the two-electron state in the alternate
transformed representation, $e^\dagger({\bf k})e^\dagger({\bf q})|0\rangle$,
represents charged fermions accompanied by the electromagnetic fields required
for them to obey Gauss's law. Nevertheless, the anticommutation rules
Eqs.~(\ref{eq:X1}) and (\ref{eq:X2}) demonstrate that $e^\dagger({\bf
k})e^\dagger({\bf q})|0\rangle = -e^\dagger({\bf q})e^\dagger({\bf
k})|0\rangle$, and that $e^\dagger({\bf k})e^\dagger({\bf k})|0\rangle=0$. The
particular form of $\tilde{\psi}({\bf x})$ given in Eq.~(\ref{eq:tildepsi})
only applies to the covariant gauges and to this method of quantization. In
other gauges, and with other methods of implementing constraints, the spinor
fields that implement Gauss's law will have a different representation, and
questions about the statistics of electron-positron states that obey Gauss's
law arise is a different way. In Sec.~\ref{sec:coulomb}, we will formulate this
theory in the Coulomb gauge and confirm the result that the charged particle
states obey standard Fermi statistics.

\section{Lorentz transformations of the photon states}
In a $(3+1)$-dimensional space, photons have two possible polarization modes;
and photons in definite helicity states transform into themselves under Lorentz
transformations. Photons share this property with all other zero-mass particles
\cite{fierz}. In contrast, massive spin-one excitations of gauge fields in
definite helicity states in one Lorentz frame, are observed as mixtures of
helicity states in other Lorentz frames. The model we are examining in this
work offers an interesting illustration of how the restriction to $2+1$
dimensions and the topological mass term affect the Lorentz transformations of
particle states. The photons in this model are massive, and propagate with
velocities $v<c$. Nevertheless, there is only a single polarization mode
available for propagating excitations that correspond to observable particles.
No second helicity mode is available with which topologically massive photons
can mix under Lorentz transformations, even though the photons are excitations
not of a scalar, but a vector field. It therefore becomes interesting to
examine how these photon states transform under a Lorentz boost.

To facilitate this investigation, we shift to a description of excitation
operators that have an invariant norm under Lorentz transformations. We
observe, for example that the norm of the one-particle state  $a^\dagger({\bf
k})|0\rangle$,
\begin{equation}
\left|a^\dagger({\bf k})|0\rangle\right|^2 = \sum_{\bf q}\langle 0|[a({\bf
q}),a^\dagger({\bf k})]|0\rangle = \int d{\bf q}\ \delta({\bf k-q}),
\end{equation}
is not a Lorentz scalar because $d{\bf k}$ is not the Lorentz invariant measure
for the phase space. The invariant measure can be established by noting that
the invariant delta function
\begin{equation}
\delta({\bf k}-{\bf q})\delta(k_0-q_0)\delta(q_\mu q^\mu - m^2)\Theta(q_0) =
\frac{\delta({\bf k}-{\bf q})\delta(k_0-\omega_q)}{2\omega_q},
\end{equation}
so that the states $A^\dagger({\bf k})|0\rangle$, created by operators that
obey
\begin{equation}
[A({\bf k}),A^\dagger({\bf q})] = 2\omega_k(2\pi)^2\delta({\bf k}-{\bf q}),
\label{eq:AAdagger}
\end{equation}
have unit norms in every Lorentz frame. The equivalently normalized ghost
operators obey
\begin{equation}
[A_Q({\bf k}),A_R^\star({\bf q})] = [A_R({\bf k}),A_Q^\star({\bf q})] =
2k(2\pi)^2\delta({\bf k}-{\bf q}).
\label{eq:ARAQdagger}
\end{equation}
Hence, the boost operator $\bar{K}_l$ for the interaction-free theory is
written as
\begin{eqnarray}
\bar{K}_l &=& \sum_{\bf k}
\frac{m\epsilon_{ln}k_n}{2k^2\omega_k}\,A^\dagger({\bf k})A({\bf k}) -
\sum_{\bf k}\frac{64k^2\epsilon_{ln}k_n}{m^4}\,A_Q^\star({\bf k})A_Q({\bf
k})\nonumber\\
\nonumber\\
&+& \sum_{\bf k}\frac{i}{4}\left[\frac{\partial}{\partial k_l}\,A^\dagger({\bf
k})A({\bf k}) - A^\dagger({\bf k})\frac{\partial}{\partial k_l}\,A({\bf
k})\right]\nonumber\\
&+& \sum_{\bf k}\frac{5ik_l}{4k^2}\left[A_Q^\star({\bf k})A_R({\bf k}) -
A_R^\star({\bf k})A_Q({\bf k})\right]\nonumber\\
&+& \sum_{\bf k} \frac{i}{2}\left[\frac{\partial}{\partial k_l}\,A_Q^\star({\bf
k})A_R({\bf k}) - A_R^\star({\bf k})\frac{\partial}{\partial k_l}\,A_Q({\bf
k})\right]\nonumber\\
&-& (1-\gamma)\sum_{\bf k}\frac{16ik^3}{m^3}\left[\frac{\partial}{\partial
k_l}\,A_Q^\star({\bf k})A_Q({\bf k}) - A_Q^\star({\bf
k})\frac{\partial}{\partial k_l}\,A_Q({\bf k})\right].
\end{eqnarray}
Using the commutations rules given by Eqs.~(\ref{eq:AAdagger}) and
(\ref{eq:ARAQdagger}), we find that
\begin{equation}
\delta A^\dagger({\bf k}) =
\left[\frac{im\epsilon_{ln}k_n}{k^2}\,A^\dagger({\bf k}) -
\omega_k\frac{\partial}{\partial k_l}\,A^\dagger({\bf k})\right]\delta\beta_l
\label{eq:delAdagger}
\end{equation}
and
\begin{equation}
\delta A_Q^\star({\bf k}) = -\left[\frac{5k_l}{2k}\,A_Q^\star({\bf k}) +
k\frac{\partial}{\partial k_l}\,A^\star_Q({\bf k})\right]\delta\beta_l,
\end{equation}
where $\delta\xi = i[\bar{K}_l,\xi]\,\delta\beta_l$.
Equation~(\ref{eq:delAdagger}) demonstrates that the particle state
$A^\dagger({\bf k})|0\rangle$ is Lorentz-transformed into itself. The phase
factor $\delta\beta_l\epsilon_{ln}k_n/k^2$ generated by the boost operator
$\bar{K}_l$, which appears in Eq.~(\ref{eq:delAdagger}), is a cocycle
\cite{cocycle}. This phase factor has no physical implications. The physically
observable consequence of Eq.~(\ref{eq:delAdagger}) is that, under a Lorentz
transformation, the topologically massive photon states behave like the
excitations of a scalar field---each photon state transforms only into itself
at a new space-time point.

\section{Formulation of the theory in the Coulomb gauge}
\label{sec:coulomb}
In the Coulomb gauge, the gauge field $A_0$ is not involved in the gauge
condition, so that a gauge-fixing term cannot be used to generate a canonical
momentum conjugate to $A_0$. The quantization procedure used in the covariant
and temporal gauge formulations of the theory therefore is not well-suited for
the Coulomb gauge. The most convenient way to quantize in the Coulomb gauge is
to use the Dirac-Bergmann (DB) procedure. In this method, the canonical
``Poisson'' commutators (anticommutators) are replaced by their respective
Dirac commutators (anticommutators), which apply to the fields that obey all
the constraints of the theory.  Since the Dirac and the canonical commutators
(anticommutators) can, and often do, differ from each other, this method
enables us to investigate whether the Dirac anticommutator for the spinor field
$\psi$ and its adjoint $\psi^\dagger$ differ from the corresponding canonical
anticommutator. A discrepancy between the Dirac and canonical anticommutators
for the spinor fields could signal the development of ``exotic'' fractional
statistics due to the imposition of Gauss's law. On the other hand, identity of
the Dirac and the canonical anticommutators for the spinor fields demonstrate
that the excitations of the charged spinor field that obey Gauss's law (as well
as all other constraints) also obey standard Fermi statistics. The question,
whether the imposition of Gauss's law produces charged particle excitations
that are subject to exotic statistics, therefore arises in a new way in the
Coulomb gauge. In this section, we will carry out this quantization procedure
and demonstrate explicitly that the implementation of Gauss's law for the
charged spinor field does not change the anticommutation rule for $\psi$ and
$\psi^\dagger$, and does not cause the excitations of these fields to develop
exotic fractional statistics.

The Lagrangian density for MCS theory in the Coulomb gauge is given by
\begin{eqnarray}
{\cal L} &=& -\case 1/4 F_{ln}F_{ln} + \case 1/2 F_{0l}F_{0l} + \case 1/4
m\epsilon_{ln}(F_{ln}A_0 + 2F_{0l}A_n)\nonumber\\
&+& j_lA_l - j_0A_0 - G\partial_lA_l  + \bar{\psi}(i\gamma^\mu\partial_\mu -
M)\psi.
\label{eq:LD}
\end{eqnarray}
This Lagrangian differs from Eq.~(\ref{eq:calL1}) only in that the gauge-fixing
term $-G\partial_\mu A^\mu$ is replaced by $-G\partial_lA_l$. We have included
a gauge-fixing term for the Coulomb gauge in Eq.~(\ref{eq:LD}) to avoid first
class constraints and to enable us to develop all the constraints
systematically from the Lagrangian. The Euler-Lagrange equations in the Coulomb
gauge are
\begin{equation}
\partial_0F_{0l} + m\epsilon_{ln}F_{0n} - \partial_nF_{ln}-\partial_lG = j_l,
\label{eq:jl}
\end{equation}
\begin{equation}
\case 1/2 m\epsilon_{ln}F_{ln} - \partial_lF_{0l} + \partial_0G = j_0,
\label{eq:Couleqofmot}
\end{equation}
\begin{equation}
\partial_lA_l = 0,
\end{equation}
and
\begin{equation}
(M - i\gamma^\mu D_\mu)\psi = 0.
\label{eq:diraceqC}
\end{equation}
The momenta conjugate to the gauge fields are $\Pi_l =  F_{0l} + \case1/2
m\epsilon_{ln}A_n$; $\Pi_0 = 0$ and $\Pi_{\text{G}}=0$, where $\Pi_0$ and
$\Pi_{\text{G}}$ are the momenta conjugate to $A_0$ and $G$, respectively. For
the spinor fields, we have $\Pi_\psi = i\psi^\dagger$ and
$\Pi_{\psi^\dagger}=0$ as the momenta conjugate to $\psi$ and $\psi^\dagger$,
respectively. We have identified the following primary constraints:
\begin{equation}
{\cal C}_1 = \Pi_0  \approx 0,
\label{eq:C1}
\end{equation}
\begin{equation}
{\cal C}_2 = \Pi_{\text{G}} \approx 0,
\label{eq:C2}
\end{equation}
\begin{equation}
{\cal C}_\psi = \Pi_\psi - i\psi^\dagger \approx 0,
\label{eq:Cpsi}
\end{equation}
and
\begin{equation}
{\cal C}_{\psi^\dagger} = \Pi_{\psi^\dagger} \approx 0.
\label{eq:Cpsidagger}
\end{equation}

The time-evolution operator is the total Coulomb gauge Hamiltonian,
$H_{\text{T}}{}^{\text{C}}$, given by
\begin{eqnarray}
H_{\text{T}}{}^{\text{C}} &=& \int d{\bf x}\ \psi^\dagger(\gamma_0M -
i\gamma_0\gamma_l\partial_l)\psi + \int d{\bf x}\ \left[\case 1/2 \Pi_l\Pi_l +
\case 1/4 F_{ln}F_{ln}+A_0\partial_l\Pi_l\right.\nonumber\\
&+& \case 1/8 m^2A_lA_l + \case 1/2 m\epsilon_{ln}A_l\Pi_n - \case 1/4
m\epsilon_{ln}F_{ln}A_0 + G\partial_lA_l + j_0A_0 - j_lA_l\nonumber\\
&-& \left.{\cal U}_1{\cal C}_1 - {\cal U}_2{\cal C}_2 - {\cal C}_\psi{\cal
U}_\psi - {\cal U}_{\psi^\dagger}{\cal C}_{\psi^\dagger}\right]
\end{eqnarray}
where ${\cal U}_1, \ldots, {\cal U}_4$ designate arbitrary functions that
commute with all operators; ${\cal U}_\psi$ and ${\cal U}_{\psi^\dagger}$
designate arbitrary functions that are Grassmann numbers, which anticommute
with all fermion fields and with Grassmann numbers, but commute with bosonic
operators and with ${\cal U}_1, \ldots, {\cal U}_4$. In the imposition of
constraints, we will use the Poisson bracket, $[\![A,B]\!]$, of two operators
$A$ and $B$, defined as $[\![A,B]\!] = AB - (-1)^{n(A) n(B)}BA$, where $n(P)$
is an index for the operators $P$;\footnote{We generally follow the conventions
in Sundermeyer \cite{sundermeyer}. The definition of Poisson bracket used here,
however, differs from Sundermeyer's definition by a factor $i$.} $n(P)=0$ if
$P$ is a bosonic operator, such as a gauge field or a bilinear combination of
fermion fields; and $n(P)=1$ if $P$ is a Grassmann number, or a fermionic
operator such as $\psi$ or $\psi^\dagger$. The Poisson bracket $[\![A,B]\!]$ is
the commutator $[A,B]$ when $A$ and $B$ are both bosonic operators, or if one
is bosonic and the other fermionic. But $[\![A,B]\!]$ is the anticommutator
$\{A,B\}$ when $A$ and $B$ are both fermionic operators.

We use the total Hamiltonian to generate the further constraints needed to
maintain the stability of the primary constraints under time evolution. For
this purpose, we evaluate time derivatives of the primary constraints by using
the equation $\partial_0{\cal C}_i = i[\![H_{\text{T}}{}^{\text{C}},C_i]\!]$,
and set $\partial_0C_i \approx 0$. In this way, we find that $\partial_0{\cal
C}_1 \approx 0$ leads to the secondary constraint ${\cal C}_3 \approx 0$ where
\begin{equation}
{\cal C}_3 = \partial_l\Pi_l - \case 1/4 m\epsilon_{ln}F_{ln} + j_0,
\end{equation}
which implements Gauss's law. $\partial_0{\cal C}_3 \approx 0$ leads to
\begin{equation}
\partial_l\partial_lG + \partial_lj_l - e\psi^\dagger{\cal U}_\psi + e{\cal
U}_{\psi^\dagger}\psi=0
\label{eq:def1}
\end{equation}
which does not generate a tertiary constraint. The stability of the constraint
$\Pi_{\text{G}} \approx 0$ is obtained by setting $\partial_0{\cal C}_2 \approx
0$ which leads to the secondary constraint ${\cal C}_4 \approx 0$ and the
tertiary constraint ${\cal C}_5 \approx 0$ where
\begin{equation}
{\cal C}_4 = \partial_lA_l
\end{equation}
and
\begin{equation}
{\cal C}_5 = \partial_l\Pi_l - \partial_l\partial_lA_0 + \case 1/4
m\epsilon_{ln}F_{ln}.
\end{equation}
The constraint ${\cal C}_4 \approx 0$ implements the gauge condition for the
Coulomb gauge, and ${\cal C}_5 \approx 0$ is required for consistency between
Eq.~(\ref{eq:Couleqofmot}) and the Coulomb gauge condition. The constraint
equation $\partial_0{\cal C}_5 \approx 0$ contains ${\cal U}_1$, thus does not
lead to any further constraints. The constraints $\partial_0{\cal
C}_\psi\approx 0$ and $\partial_0{\cal C}_{\psi^\dagger} \approx 0$ result in
the following expressions for ${\cal U}_\psi$ and ${\cal U}_{\psi^\dagger}$:
\begin{equation}
{\cal U}_\psi = ieA_0\psi -ieA_l\gamma_l\psi + iM\gamma_0\psi +
\gamma_0\gamma_l\partial_l\psi
\label{eq:Upsi}
\end{equation}
and
\begin{equation}
{\cal U}_{\psi^\dagger} = ieA_0\psi^\dagger - ieA_l\psi^\dagger\gamma_l +
iM\psi^\dagger\gamma_0 - \partial_l\psi^\dagger\gamma_0\gamma_l.
\label{eq:Upsidagger}
\end{equation}
Substitution of Eqs.~(\ref{eq:Upsi}) and (\ref{eq:Upsidagger}) into
Eq.~(\ref{eq:def1}) yields still another constraint, ${\cal C}_6 \approx 0$,
where
\begin{equation}
{\cal C}_6 = \partial_l\partial_lG
\end{equation}
which is necessary for consistency between Eq.~(\ref{eq:jl}) and Gauss's law.
The constraint $\partial_0{\cal C}_6 \approx 0$ is an equation containing
${\cal U}_2$ but does not lead to any further constraint.

The preceding analysis leads to eight second-class constraints for this gauge
theory. Imposition of the constraints requires that we form the matrix ${\cal
M}({\bf x},{\bf y})$, whose elements are ${\cal M}_{ij}({\bf x},{\bf
y})=[\![{\cal C}_i({\bf x}),{\cal C}_j({\bf y})]\!]$. We assign the values
${\cal C}_1, \ldots, {\cal C}_{8}$ to the descending horizontal rows of the
matrix, as well as to the sequence of vertical columns, where ${\cal C}_1,
\ldots, {\cal C}_6$ refer to the previously defined constraints; for simplicity
we will designate ${\cal C}_\psi$ and ${\cal C}_{\psi^\dagger}$ as ${\cal C}_7$
and ${\cal C}_{8}$, respectively. The matrix ${\cal M}({\bf x},{\bf y})$ is
given by
\begin{equation}
{\cal M}({\bf x},{\bf y}) = \left(\begin{array}{cccccccc}
0	&	0	&	0	&	0	&i\nabla^2&	0	&	0	&	0\\[10pt]
0	&	0	&	0	&	0	&	0	&-i\nabla^2&	0	&	0\\[10pt]
0	&	0	&	0	&i\nabla^2&	0	&	0	&ie\psi^\dagger({\bf x})&-ie\psi({\bf x})\\[10pt]
0	&	0	&-i\nabla^2&	0	&-i\nabla^2&	0	&	0	&	0\\[10pt]
-i\nabla^2&	0	&	0	&i\nabla^2&	0	&	0	&	0	&	0\\[10pt]
0	&i\nabla^2&	0	&	0	&	0	&	0	&	0	&	0\\[10pt]
0	&	0	&-ie\psi^\dagger({\bf x})&	0	&	0	&	0	&	0	&	1\\[10pt]
0	&	0	&ie\psi({\bf x})&	0	&	0	&	0	&	1	&	0\end{array}
\right)\delta({\bf x-y}).
\end{equation}
The matrix ${\cal M}({\bf x},{\bf y})$ has an inverse, ${\cal Y}({\bf x},{\bf
y})$, given by
\begin{equation}
{\cal Y}({\bf x},{\bf y}) = \left(\begin{array}{cccccccc}
0 & 0 & \displaystyle\frac{-i}{\nabla^2} & 0 & \displaystyle\frac{i}{\nabla^2}
& 0 & e\psi({\bf y})\displaystyle\frac{1}{\nabla^2} & -e\psi^\dagger({\bf
y})\displaystyle\frac{1}{\nabla^2}\\[10pt]
0 & 0 & 0 & 0 & 0 & \displaystyle\frac{-i}{\nabla^2} & 0 & 0\\[10pt]
\displaystyle\frac{i}{\nabla^2} & 0 & 0 & \displaystyle\frac{i}{\nabla^2} & 0 &
0 & 0 & 0\\[10pt]
0 & 0 & \displaystyle\frac{-i}{\nabla^2} & 0 & 0 & 0 & e\psi({\bf
y})\displaystyle\frac{1}{\nabla^2} & -e\psi^\dagger({\bf
y})\displaystyle\frac{1}{\nabla^2}\\[10pt]
\displaystyle\frac{-i}{\nabla^2} & 0 & 0 & 0 & 0 & 0 & 0 & 0\\[10pt]
0 & \displaystyle\frac{-i}{\nabla^2} & 0 & 0 & 0 & 0 & 0 & 0\\[10pt]
e\psi({\bf y})\displaystyle\frac{1}{\nabla^2} & 0 & 0 & e\psi({\bf
y})\displaystyle\frac{1}{\nabla^2} & 0 & 0 & 0 & 1\\[10pt]
-e\psi^\dagger({\bf y})\displaystyle\frac{1}{\nabla^2} & 0 & 0 &
-e\psi^\dagger({\bf y})\displaystyle\frac{1}{\nabla^2} & 0 & 0 & 1 & 0
 \end{array}\right)\!\delta({\bf x-y}).
\end{equation}
We note that
\begin{equation}
\int d{\bf z}\ {\cal M}_{ik}({\bf x},{\bf z}){\cal Y}_{kj}({\bf z},{\bf
y})=\int d{\bf z}\ {\cal Y}_{ik}({\bf x},{\bf z}){\cal M}_{kj}({\bf z},{\bf y})
= \delta_{ij}\delta({\bf x}-{\bf y}).
\end{equation}
We apply
\begin{equation}
[\![\xi({\bf x}),\zeta({\bf y})]\!]^{\text{D}} = [\![\xi({\bf x}),\zeta({\bf
y})]\!] - \sum_{i,j=1}^{8}\int d{\bf z}\,d{\bf z}^\prime\ [\![\xi({\bf
x}),{\cal C}_i({\bf z})]\!]{\cal Y}_{ij}({\bf z},{\bf z}^\prime)[\![{\cal
C}_j({\bf z}^\prime),\zeta({\bf y})]\!]
\end{equation}
to find Dirac commutators (anticommutators) for the gauge and/or spinor fields
represented by $\xi$ and $\zeta$, and observe that these are given by
\begin{equation}
[\![\psi({\bf x}),\psi^\dagger({\bf y})]\!]^{\text{D}} = \{\psi({\bf
x}),\psi^\dagger({\bf y})\}=\delta({\bf x}-{\bf y}),
\label{eq:psipsidaggerD}
\end{equation}
\begin{equation}
[\![\psi({\bf x}),\psi({\bf y})]\!]^{\text{D}} = \{\psi({\bf x}),\psi({\bf
y})\}=0,
\label{eq:psipsiD}
\end{equation}
\begin{equation}
[\![A_l({\bf x}),\Pi_n({\bf y})]\!]^{\text{D}} = i\left(\delta_{ln} -
\frac{\partial_l\partial_n}{\nabla^2}\right)\delta({\bf x-y}),
\end{equation}
\begin{equation}
[\![A_0({\bf x}),A_l({\bf y})]\!]^{\text{D}}=[\![A_l({\bf x}),A_n({\bf
y})]\!]^{\text{D}}=[\![A_l({\bf x}),\psi({\bf y})]\!]^{\text{D}} = 0,
\end{equation}
and
\begin{equation}
[\![A_0({\bf x}),\psi({\bf y})]\!]^{\text{D}} = e\psi({\bf
y})\frac{1}{\nabla^2}\,\delta({\bf x-y}).
\end{equation}

Equations~(\ref{eq:psipsidaggerD}) and (\ref{eq:psipsiD}) demonstrate that the
constrained spinor field obeys standard anticommutation rules, and not a graded
anticommutator algebra; and that the charged excitations of that spinor field
are subject to standard Fermi statistics, and not the exotic fractional
statistics that would result from a graded anticommutator algebra. In contrast
to the spinor field, the Dirac commutators of the gauge fields differ
substantially both from the unconstrained canonical commutators, and also from
their corresponding values in the temporal gauge. The observation that the
spinor anticommutation rule is unaffected by constraints, and identical in the
Coulomb and temporal gauges, therefore is not trivial.

The constrained Hamiltonian in the Coulomb gauge which now incorporates all the
constraints ${\cal C}_1$, \ldots, ${\cal C}_8$, is found to be
\begin{equation}
H_{\text{C}} = \int d{\bf x}\ \left[\case 1/2 \Pi_l^{\text{T}}\Pi_l^{\text{T}}
+ \case 1/4 F_{ln}F_{ln} + \case 1/2 m^2 A_l^{\text{T}}A_l^{\text{T}} -
m\epsilon_{ln}\partial_lA_n^{\text{T}}\nabla^{-2}j_0 - \case 1/2
j_0\nabla^{-2}j_0 - j_lA_l^{\text{T}}\right]
\end{equation}
where $A_l^{\text{T}}$ and $\Pi_l^{\text{T}}$ are the transverse components of
$A_l$ and $\Pi_l$, respectively; they are given by
\begin{equation}
A_l^{\text{T}} = \left(\delta_{ln} -
\frac{\partial_l\partial_n}{\nabla^2}\right)A_n
\end{equation}
and
\begin{equation}
\Pi_l^{\text{T}} = \left(\delta_{ln} -
\frac{\partial_l\partial_n}{\nabla^2}\right)\Pi_n.
\end{equation}
We need to find a suitable representation for the gauge fields in terms of
particle creation and annihilation operators, in the Coulomb gauge formulation,
just as we did in Sec.~\ref{sec:formulation}, when developing this model in the
covariant gauges. The criterion of suitability is similar: The Hamiltonian for
the free, noninteracting system of photons and electrons must have the
perturbative vacuum $|0\rangle$, and the single-particle photon and
electron-positron states, $a^\dagger({\bf k})|0\rangle$, $e^\dagger({\bf
k})|0\rangle$ and $\bar{e}^\dagger({\bf k})|0\rangle$ respectively, as
eigenstates, so that the interaction-free time-evolution operator $e^{-iH_0t}$
propagates these and other multi-particle states without changing their form or
particle content. Since in this formulation the Dirac-Bergmann procedure has
been implemented, time-evolution is restricted to the constraint surface, and
all constraints (including Gauss's law) apply. In the case of the Coulomb
gauge, only the transverse gauge fields have commutation rules that need to be
accommodated, so that ghost excitations of the gauge field are not required.
The electrons and positrons are represented precisely as in the other gauges.

A suitable representation of the gauge fields can be based on the transverse
part of $A_l({\bf x})$ in the covariant gauge, and leads to
\begin{equation}
A_l{}^{\text{T}}({\bf x}) = \sum_{\bf
k}\frac{i\epsilon_{ln}k_n}{k\sqrt{2\omega_k}}\left[a({\bf k})e^{i{\bf k\cdot
x}}- a^\dagger({\bf k})e^{-i{\bf k\cdot x}}\right],
\label{eq:tildeA}
\end{equation}
\begin{equation}
\Pi_l{}^{\text{T}}({\bf x}) = \sum_{\bf
k}\frac{\sqrt{\omega_k}\epsilon_{ln}k_n}{\sqrt{2}k}\left[a({\bf k})e^{i{\bf
k\cdot x}} + a^\dagger({\bf k})e^{-i{\bf k\cdot x}}\right],
\label{eq:tildePi}
\end{equation}
and
\begin{equation}
A_0{}^{\text{D}}({\bf x}) = \sum_{\bf k}\frac{m}{k\sqrt{2\omega_k}}\left[a({\bf
k})e^{i{\bf k\cdot x}}+ a^\dagger({\bf k})e^{-i{\bf k\cdot x}}\right] +
\sum_{\bf k}\frac{1}{k^2}\,j_0({\bf k})e^{i{\bf k\cdot x}}.
\label{eq:tildeA0}
\end{equation}
The Hamiltonian $H_{\text{C}}$ then becomes
\begin{eqnarray}
H_{\text{C}} &=& \sum_{\bf k}\omega_ka^\dagger({\bf k})a({\bf k}) + \sum_{\bf
k}\frac{1}{2k^2}\,j_0({\bf k})j_0(-{\bf k})\nonumber\\
&+&\sum_{\bf k}\frac{m}{k\sqrt{2\omega_k}}\left[j_0(-{\bf k})a({\bf
k})+j_0({\bf k})a^\dagger({\bf k})\right]\nonumber\\
&+& \sum_{\bf k}\frac{i\epsilon_{ln}k_l}{k\sqrt{2\omega_k}}\left[j_n(-{\bf
k})a({\bf k}) - j_n({\bf k})a^\dagger({\bf k})\right] + H_{\bar{e}e}.
\label{eq:tildeh}
\end{eqnarray}
Although the representations of $A_l^{\text{T}}({\bf x})$,
$\Pi_l^{\text{T}}({\bf x})$ and $A_0({\bf x})$ given in
Eqs.~(\ref{eq:tildeA})--(\ref{eq:tildeA0}) are suitable, they are not unique.
Other suitable representations would satisfy the requirements specified above
equally well. We therefore cannot assume that the Hilbert space in which these
operators---and the representation of $H_{\text{C}}$ given in
Eq.~(\ref{eq:tildeh})---operate, is identical to the one previously established
for the corresponding operators in the covariant gauges. All the requirements
we stipulated for representations to be suitable remain unaffected by
similarity transformations---i.e., unitary transformations carried out on the
operators as well as states of a particular representation. It is therefore
necessary to consider the possibility that the appropriate Fock space for this
representation of MCS theory in the Coulomb gauge differs from the Fock space
$\{|N\rangle\}$ we used for the quotient space of the covariant gauge
formulation of this model, by just such a similarity transformation. In fact,
this possibility becomes a virtual certainty, in view of the fact that
$H_{\text{C}}$ does not have the identical form as $\hat{H}_{\text{quot}}$, the
generator of time displacements in the Fock space $\{|N\rangle\}$ of the
covariant gauge formulation developed in Secs.~\ref{sec:formulation} and
\ref{sec:implofgausslaw}. As in the case of $(3+1)$-dimensional QED in axial
gauges \cite{el3}, the reason for the discrepancy in form between the two
Hamiltonians, $H_{\text{C}}$ and $\hat{H}_{\text{quot}}$, lies in the fact that
the two operate in different Fock spaces. We can expect that the Fock space on
which the operators in Eqs.~(\ref{eq:tildeA})--(\ref{eq:tildeh}) act, consists
of state vectors $|\phi_i\rangle = e^\Gamma|N_i\rangle$, where $|N_i\rangle$
designates the state vectors in the Fock space whose single-particle elements
are the familiar $a^\dagger({\bf k})|0\rangle$, $e^\dagger({\bf k})|0\rangle$
and $\bar{e}^\dagger({\bf k})|0\rangle$. We can then proceed to carry out a
similarity transformation in which all the operators transform according to
$\hat{\cal P}=e^\Gamma{\cal P}e^{-\Gamma}$; the similarly transformed states
then are $|\hat{\phi}_i\rangle = e^{-\Gamma}|\phi_i\rangle$. If the objective
of this transformation has been met, then we will find that
$|\hat{\phi}_i\rangle=|N_i\rangle$, so that the transformed states will be the
elements of the Fock space $\{|N\rangle\}$ familiar to us as the quotient space
of the covariant gauge formulation. To carry out such a program, we choose
\begin{equation}
\Gamma = -\sum_{\bf k}\frac{m}{\sqrt{2}k\omega_k^{3/2}}\left[a({\bf
k})j_0(-{\bf k}) - a^\dagger({\bf k})j_0({\bf k})\right],
\end{equation}
and find the transformed operators
\begin{eqnarray}
\hat{H}_{\text{C}} &=& \sum_{\bf k}\omega_ka^\dagger({\bf k})a({\bf k}) +
\sum_{\bf k}\frac{1}{2\omega_k^2}\,j_0({\bf k})j_0(-{\bf k}) + \sum_{\bf
k}\frac{im\epsilon_{ln}k_l}{k^2\omega_k^2}\,j_n({\bf k})j_0(-{\bf
k})\nonumber\\
&+&\sum_{\bf k}\frac{mk_l}{\sqrt{2}k\omega_k^{3/2}}\left[j_l(-{\bf k})a({\bf
k})+j_l({\bf k})a^\dagger({\bf k})\right]\nonumber\\
&+& \sum_{\bf k}\frac{i\epsilon_{ln}k_l}{k\sqrt{2\omega_k}}\left[j_n(-{\bf
k})a({\bf k}) - j_n({\bf k})a^\dagger({\bf k})\right] + H_{\bar{e}e},
\label{eq:hath}
\end{eqnarray}
\begin{equation}
\hat{A}_l{}^{\text{T}}({\bf x}) = \sum_{\bf
k}\frac{i\epsilon_{ln}k_n}{k\sqrt{2\omega_k}}\left[a({\bf k})e^{i{\bf k\cdot
x}}- a^\dagger({\bf k})e^{-i{\bf k\cdot x}}\right] - \sum_{\bf k}
\frac{im\epsilon_{ln}k_n}{k^2\omega_k^2}\,j_0({\bf k})e^{i{\bf k\cdot x}},
\label{eq:hatA}
\end{equation}
\begin{equation}
\hat{\Pi}_l{}^{\text{T}}({\bf x}) = \sum_{\bf
k}\frac{\sqrt{\omega_k}\epsilon_{ln}k_n}{\sqrt{2}k}\left[a({\bf k})e^{i{\bf
k\cdot x}} + a^\dagger({\bf k})e^{-i{\bf k\cdot x}}\right]
\label{eq:hatPi}
\end{equation}
and
\begin{equation}
\hat{A}_0{}^{\text{D}}({\bf x}) = \sum_{\bf
k}\frac{m}{k\sqrt{2\omega_k}}\left[a({\bf k})e^{i{\bf k\cdot x}}+
a^\dagger({\bf k})e^{-i{\bf k\cdot x}}\right] + \sum_{\bf
k}\frac{1}{\omega_k^2}\,j_0({\bf k})e^{i{\bf k\cdot x}}.
\label{eq:hatA0}
\end{equation}

The Fock space $\{|N\rangle\}$ is the appropriate Hilbert space for the Coulomb
gauge formulation {\em after} the similarity transformation $\hat{\cal
P}=e^\Gamma{\cal P}e^{-\Gamma}$ and
$|n_i\rangle=|\hat{\phi}_i\rangle=e^{-\Gamma}|\phi\rangle$ has been carried
out. $\hat{H}_{\text{C}}$ operates in the same Fock space as does
$\hat{H}_{\text{quot}}$, and the two operators have the identical form. State
vectors in the Fock space $\{|N\rangle\}$, representing systems of electrons,
positrons and the topologically massive, propagating excitations of the gauge
field, are time translated by the same time evolution operator in both the
covariant and the Coulomb gauges. Earlier work demonstrated that the same
time-evolution operator also time translates these state vectors in the
temporal gauge \cite{hl}.

It is apparent from the preceding discussion that, had we initially chosen Eqs.
(\ref{eq:hath})--(\ref{eq:hatA0}) to represent $A_l^{\text{T}}({\bf x})$,
$\Pi_l^{\text{T}}({\bf x})$ and $A_0({\bf x})$, we would have immediately
obtained the desired form $\hat{H}_{\text{C}}$ for the Hamiltonian for MCS
theory in the Coulomb gauge, and would have had no occasion to carry out any
unitary transformations. However, since there is no systematic way of initially
recognizing the appropriate representation of $A_l^{\text{T}}({\bf x})$,
$\Pi_l^{\text{T}}({\bf x})$ and $A_0({\bf x})$ that leads to this desired form
for the Hamiltonian, we have deliberately avoided making the most convenient
choice of representation from the start. It is important to formulate the
question, whether two different representations describe the same physical
system, in terms of the identity of two equivalence classes, in which the
operators and states that are members of a class are related by similarity
transformations. It is not sufficient, in testing whether operators,
constructed with randomly chosen representations of space-time fields, have the
same form. This point has been discussed in greater detail elsewhere \cite{el3}
but applies here as well.

\section{Is CS theory the large {\bf\it \lowercase{m}} limit of MCS theory?}
The Lagrangians for CS theory and MCS theory differ only by the Maxwell kinetic
energy term, which is included in the latter but absent from the former. The
relative size of the CS term and the Maxwell kinetic energy term is tuned by
the CS coupling constant $m$, and in the limit $m\rightarrow\infty$ the Maxwell
kinetic energy term becomes vanishingly small relative to the CS term. The
question therefore naturally arises, whether CS theory is approached as a
well-defined limit of MCS theory as $m\rightarrow\infty$. The results obtained
in this work provide some insights into that question.

The comparison between CS and MCS theory can be best approached through what we
have called the ``quotient space'' Hamiltonian, $H_{\text{quot}}$, for MCS
theory, given in  Eqs.~(\ref{eq:Hhat0}) and (\ref{eq:hatHI}). $H_{\text{quot}}$
is the form the Hamiltonian takes in the quotient space for the Fock space
$\{|n\rangle\}$, in the representation in which the latter implements Gauss's
law and the gauge choice; and $H_{\text{quot}}$ has the same form in covariant,
temporal and Coulomb gauges. Ambiguities in $H_{\text{quot}}$, of the form
$h=i[H_{\text{quot}},\chi]=i[H_0,\chi]$, can arise; in
Sec.~\ref{sec:implofgausslaw}, we have discussed such ambiguous terms, and have
shown that they cannot affect the $S$-matrix, and that they can be transformed
away by unitary transformations. We will assume here that such terms have been
transformed away, and are not included in $H_{\text{quot}}$. Apart from such
ambiguities, $H_{\text{quot}}$ consists of a ``free'' part that counts the
kinetic energy of propagating massive photons and electrons $(e^-$ and $e^+)$;
interaction terms
\begin{equation}
{\sf H}_{\text{a}} = \int d{\bf x}\,d{\bf y}\ j_0({\bf x})j_0({\bf
y})K_0(m|{\bf x-y}|)
\end{equation}
and
\begin{equation}
{\sf H}_{\text{b}} = \int d{\bf x}\,d{\bf y}\ j_0({\bf x})\epsilon_{ln}j_l({\bf
y})(x-y)_n{\cal F}(|{\bf x-y}|)
\end{equation}
which describe nonlocal interactions between charges, and between charges and
transverse currents, respectively; and finally, parts of $H_{\text{quot}}$
describe interactions between the massive propagating photons and electrons. In
the limit $m\rightarrow\infty$, the following observations can be made about
the component parts of $H_{\text{quot}}$: ${\sf H}_{\text{a}}$ vanishes in that
limit, and its leading term in powers of $1/m$ is of order $1/m^2$. The
modified Bessel function $K_0(\xi)$ that appears in ${\sf H}_{\text{a}}$ takes
the asymptotic form
$$
K_0(\xi) \rightarrow \sqrt{\frac{\pi}{2\xi}}\,e^{-\xi}
$$
in the limit $\xi \rightarrow \infty$. And for $\xi = m\,|{\bf x-y}|$, $\xi
\rightarrow \infty$ as $m \rightarrow \infty$ for all values of $|{\bf x-y}|$
except $|{\bf x-y}|=0$; $j_0({\bf x})$ and $j_0({\bf y})$ are operators whose
matrix elements will be superpositions of products of nonsingular wavefunctions
given in Eqs.~(\ref{eq:ubfk}) and (\ref{eq:vbfk}) or (\ref{eq:u+}) and
(\ref{eq:u-}). The integration $\int d{\bf x}\,d{\bf y}\ \cdots$ can be
transformed to $\int d{\bf r}\,d\mbox{\boldmath $\rho$}\ \cdots$, where ${\bf
r}={\bf x-y}$ and $\mbox{\boldmath $\rho$}=\case 1/2 ({\bf x+y})$, and the
$r\,dr$ in $d{\bf r}$ regularizes the logarithmic singularity of $K_0(mr)$ at
$r=0$, so that the integrand in ${\sf H}_{\text{a}}$ vanishes at $r=0$ for all
well-behaved charge densities. ${\sf H}_{\text{a}}$ can be expressed as
\begin{equation}
{\sf H}_{\text{a}} = \int dr\ K_0(mr)f(r)
\label{eq:sfHa}
\end{equation}
with
\begin{equation}
f(r) = r\int d\Omega_{\bf r}\,d\mbox{\boldmath $\rho$}\ j_0(\mbox{\boldmath
$\rho$} + \case 1/2 {\bf r})j_0(\mbox{\boldmath $\rho$} - \case 1/2 {\bf r}).
\end{equation}
As $m\rightarrow\infty$, the integrand of Eq.~(\ref{eq:sfHa}) becomes
vanishingly small except when $r\ll1$, and in that region $f(r)$ can be
represented as a series, whose leading term makes a contribution to ${\sf
H}_{\text{a}}$ given by
\begin{equation}
{\sf H}_{\text{a}} = \lambda\int r\,dr\ K_0(mr)
\end{equation}
where
\begin{equation}
\lambda = 4\pi\int d\mbox{\boldmath $\rho$}\ [j_0(\rho)]^2.
\end{equation}
Since
\begin{equation}
\int r\,dr\ K_0(mr) = \frac{1}{m^2},
\end{equation}
this leading term of ${\sf H}_{\text{a}}$ is of order $1/m^2$. Expansions
beyond this leading order, which reflect the nonnegligible ${\bf r}$-dependence
of $j_0(\mbox{\boldmath $\rho$}\pm\case 1/2 {\bf r})$ as $m$ gets smaller, will
produce additional terms of order $(1/m)^N$ with $N>2$.

The $m$-dependence of ${\sf H}_{\text{b}}$ in the $m\rightarrow\infty$ limit is
exactly $1/m$, so that the ratio ${\sf H}_{\text{a}}/{\sf
H}_{\text{b}}\rightarrow0$ as $m\rightarrow\infty$. Moreover, in the
$m\rightarrow\infty$ limit, ${\sf H}_{\text{b}}$ approaches the expression for
the interaction between charges and transverse currents in CS theory. In CS
theory, the function ${\cal F}(mr)$ that appears in ${\sf H}_{\text{b}}$ is
replaced by the integral
$$
\frac{-1}{2\pi m|{\bf x-y}|^2}\int_0^\infty du\ J_1(u).
$$
Since $\int_0^\infty du\ J_1(u) =1$, this agrees with the large $m$ limit of
${\cal F}(mr)$ given in Eq.~(\ref{eq:calf(r)}). In the $m\rightarrow\infty$
limit, the sum of the two interactions ${\sf H}_{\text{a}} + {\sf
H}_{\text{b}}$ therefore can be seen to approach the same limit as ${\sf
H}_{\text{b}}$ alone; and that limit is the nonlocal interaction between
charges and transverse currents in CS theory.

The interactions between propagating massive photons and currents that arise in
MCS theory have no corresponding counterpart in CS theory. The massive photons
of MCS theory never disappear as $m\rightarrow\infty$. They can transmit an
interaction between charges which we will examine for the case of
electron-electron scattering. To lowest order in $1/m$, the part of the
$S$-matrix element for $e({\bf P}) + e({\bf Q}) \rightarrow e({\bf P}^\prime) +
e({\bf Q}^\prime)$ that originates from the exchange of a propagating massive
photon between electrons is given by
\begin{eqnarray}
S_{fi}^{(2)} &=& -\frac{i(2\pi)^3M^2e^2\delta^3({P + Q - P^\prime -
Q^\prime})}{\sqrt{\rule{0pt}{8pt}\bar{\omega}_{P^\prime}
\bar{\omega}_P\bar{\omega}_{Q^\prime}\bar{\omega}_Q}}\times\nonumber\\
&&\left[\frac{im\epsilon_{ll^\prime}(\bar{\omega}_{P^\prime} -
\bar{\omega}_Q)\bar{u}({\bf Q}^\prime)\gamma_{l^\prime} u({\bf P})\bar{u}({\bf
P}^\prime)\gamma_lu({\bf Q})}{\omega_{P^\prime-Q}^2[\omega_{P^\prime-Q}^2 -
(\bar{\omega}_{P^\prime} - \bar{\omega}_Q)^2 - i\epsilon]}\right.\nonumber\\
&-&\frac{im\epsilon_{ll^\prime}(\bar{\omega}_{P^\prime} -
\bar{\omega}_P)\bar{u}({\bf Q}^\prime)\gamma_{l^\prime} u({\bf P})\bar{u}({\bf
P}^\prime)\gamma_lu({\bf Q})}{\omega_{P^\prime-P}^2[\omega_{P^\prime-P}^2 -
(\bar{\omega}_{P^\prime} - \bar{\omega}_P)^2 - i\epsilon]}\nonumber\\
&-&\frac{\bar{u}({\bf Q}^\prime)\gamma_{l} u({\bf P})\bar{u}({\bf
P}^\prime)\gamma_lu({\bf Q})}{\omega_{P^\prime-Q}^2 -
(\bar{\omega}_{P^\prime}-\bar{\omega}_Q)^2 - i\epsilon}+\frac{\bar{u}({\bf
Q}^\prime)\gamma_{l} u({\bf Q})\bar{u}({\bf P}^\prime)\gamma_lu({\bf
P})}{\omega_{P^\prime-P}^2 - (\bar{\omega}_{P^\prime}-\bar{\omega}_P)^2 -
i\epsilon}\nonumber\\
&+&\frac{(\bar{\omega}_{P^\prime}-\bar{\omega}_Q)^2u^\dagger({\bf
Q}^\prime)u({\bf P})u^\dagger({\bf P}^\prime)u({\bf
Q})}{\omega_{P^\prime-Q}^2[\omega_{P^\prime-Q}^2 -
(\bar{\omega}_{P^\prime}-\bar{\omega}_Q)^2 - i\epsilon]}\nonumber\\
&-&\left.\frac{(\bar{\omega}_{P^\prime}-\bar{\omega}_P)^2u^\dagger({\bf
Q}^\prime)u({\bf Q})u^\dagger({\bf P}^\prime)u({\bf
P})}{\omega_{P^\prime-P}^2[\omega_{P^\prime-P}^2 -
(\bar{\omega}_{P^\prime}-\bar{\omega}_P)^2 - i\epsilon]}\right].
\end{eqnarray}
The leading term in $1/m$ of $S_{fi}^{(2)}$ can be seen to be of order $1/m^2$,
so that the interaction between charged particles mediated by photon exchange
vanishes as quickly as ${\sf H}_{\text{a}}$, as $m\rightarrow\infty$, namely
one power of $1/m$ more rapidly that does the dominant interaction term, ${\sf
H}_{\text{b}}$. Photon exchange therefore will not prevent the interactions
between charged particles in MCS theory from approaching the corresponding
interaction in CS theory.

The interactions between propagating massive photons and currents also describe
electron-photon scattering, and processes in which charged particles radiate
energy in the form of massive photons. Since these processes do not, and indeed
cannot occur in CS theory, we must take account of the fact that they do not
vanish in the $m\rightarrow\infty$ limit of MCS theory. A large photon mass
does not disqualify the photon from being part of an initial state in a
scattering process. Nor is the matrix element for photon production very
sharply attenuated in the $m\rightarrow\infty$ limit. However, in that case,
the energy required to produce even a single photon increases with $m$, so that
for ordinary energy regime this process is a not a realistic option as
$m\rightarrow\infty$. Nevertheless, the interaction that produces photons in
MCS theory does not vanish in the large $m$ limit. In that sense MCS theory
never fully approaches CS theory as $m\rightarrow\infty$.

\acknowledgements
This research was supported by the Department of Energy under Grant
No.~DE-FG02-92ER40716.00.

\appendix
\section{}
In this appendix, we return to the Poincar\'e algebra---in particular to
Eq.~(\ref{eq:KlKn})---to discuss an ambiguity that arises when the same
equation is examined after the momentum representation of the gauge
fields---Eqs.~(\ref{eq:Ail})--(\ref{eq:G})---have been substituted into the
boost operators. We observe, in that case, that besides the $x_l$ in $\int
d{\bf x}\ x_l{\cal P}_0({\bf x})$ (and the corresponding $y_n$ in $\int d{\bf
y}\ y_n{\cal P}_0({\bf y})$, the only other explicit dependence on
configuration space variables is through the exponentials $e^{\pm i{\bf k\cdot
x}}$ and $e^{\pm i{\bf k\cdot y}}$ that appear in the gauge fields. The
representations of $x_l$ and $y_n$ required to facilitate the evaluation of
$K_l$ and $K_n$ are $x_l = \mp\displaystyle\frac{\partial}{\partial k_l}e^{\pm
i{\bf k\cdot x}}$ and $y_n = \mp\displaystyle\frac{\partial}{\partial
k_n}e^{\pm i{\bf k\cdot y}}$. In order to generate Dirac delta functions in
momentum space variables, and to permit the integration over these delta
functions, all partial derivatives of the form
$\displaystyle\frac{\partial}{\partial k_l}$ (in various different momentum
variables) must be integrated by parts. This step produces the derivatives of
photon creation and annihilation operators required for the angular momentum
operator
\begin{eqnarray}
J &=& -\sum_{\bf k}\frac{i\epsilon_{ln}k_n}{2} \left[\frac{\partial}{\partial
k_l}\,a^\dagger({\bf k})a({\bf k}) - a^\dagger({\bf k})\frac{\partial}{\partial
k_l}\,a({\bf k})\right]\nonumber\\
&-& \sum_{\bf k}i\epsilon_{ln}k_n\left[\frac{\partial}{\partial
k_l}\,a_Q^\star({\bf k})a_R({\bf k}) - a_R^\star({\bf
k})\frac{\partial}{\partial k_l}\,a_Q({\bf k})\right]
\label{eq:J1}
\end{eqnarray}
in Eq.~(\ref{eq:KlKn}); in addition, this process generates further partial
derivatives of the coefficients of the expressions $a({\bf k})e^{i{\bf k\cdot
x}}$ and $a^\dagger({\bf k})e^{-i{\bf k\cdot x}}$ that appear in
Eqs.~(\ref{eq:Ail})--(\ref{eq:G}). When partial derivatives of singular
coefficients arise---those for which the identities
\begin{equation}
\frac{\partial}{\partial k_l}\,\frac{k_n}{k^2} = \left[\pi\delta({\bf k}) +
\frac{1}{k^2}\right]\delta_{ln} - \frac{2k_lk_n}{k^4}
\label{eq:result1}
\end{equation}
and
\begin{equation}
\frac{\partial}{\partial k_l}\frac{k_n}{k}=\left[\pi k\delta({\bf k}) +
\frac{1}{k}\right]\delta_{ln} - \frac{k_lk_n}{k^3}
\label{eq:result2}
\end{equation}
hold---the delta functions $\delta({\bf k})$ that appear in these expressions
give rise to spurious contributions that produce an extra term, beyond
$-i\epsilon_{ln}J$ in Eq.~(\ref{eq:KlKn}). For the representation we have given
in Eqs.~(\ref{eq:Ail})--(\ref{eq:G}), the commutator $[K_l,K_n]$, evaluated in
the momentum space representation, is given by
\begin{equation}
[K_l,K_n] = -i\epsilon_{ln}J + \frac{1}{2\pi}\,im^2\epsilon_{ln}a^\dagger({\bf
0})a({\bf 0}).
\end{equation}
This discrepancy was noticed and discussed by Deser, Jackiw and Templeton
\cite{djt}. As follows from Ref.~\cite{djt}, the unitary transformation
$e^{i\phi}\,[\cdots]e^{-i\phi}$ with $\phi=-\sum_{\bf k}(\tan^{-1}
k_2/k_1)a^\dagger({\bf k})a({\bf k})$ removes the singularity in the
coefficients of $a({\bf k})e^{i{\bf k\cdot x}}$ and $a^\dagger({\bf k})e^{i{\bf
k\cdot x}}$ in Eqs.~(\ref{eq:Ail})--(\ref{eq:G}), and with it also this
discrepancy. The fact that such a discrepancy appears with singular
coefficients, points to a mathematical inconsistency that arises when we carry
out the formal manipulations required to evaluate $[K_l,K_n]$ in the momentum
representation of the gauge fields. It is not surprising, perhaps, that such
problems arise when orders of integration of operator-valued integrands are
reversed, and when Dirac delta functions are treated as functions, and not as
distributions, when integrations by parts are carried out. However, as was
pointed out in Ref.~\cite{djt}, since unitary transformations like
$e^{i\phi}\,[\cdots]e^{-i\phi}$ are able to remove this ``zero-momentum
discrepancy,'' we are dealing with a mathematical ambiguity rather than a
threat to the consistency of the Poincar\'e algebra, and with it to the
consistency of this formulation.

For completeness, we include one term in the evaluation of $[K_l,K_n]$, when
the momentum space representations of the gauge field are used---the commutator
$X_{ln}$ obtained from
\begin{equation}
X_{ln} = [\case 1/2\int d{\bf x}\ x_l\Pi_r({\bf x})\Pi_r({\bf x}), \case 1/2
m\epsilon_{ij}\int d{\bf y}\ y_nA_i({\bf y})\Pi_j({\bf y})],
\end{equation}
and given by
\begin{equation}
X_{ln} = \int d{\bf x}\,d{\bf y}\ x_ly_n \case{1}{2} im\epsilon_{ij}\Pi_i({\bf
x})\Pi_j({\bf y})\delta({\bf x-y}).
\end{equation}
In the configuration space representation, $X_{ln}$ can easily be shown to be
given by
\begin{equation}
X_{ln} = \case 1/2 im\epsilon_{ij}\int d{\bf x}\ x_lx_n\Pi_i({\bf x})\Pi_j({\bf
x}),
\label{eq:xln0}
\end{equation}
which vanishes trivially because of the antisymmetry of
$\epsilon_{ij}\Pi_i({\bf x})\Pi_j({\bf x})$. When the momentum space
representations of the gauge fields are used, however, $X_{ln}$ has the form
\begin{eqnarray}
X_{ln} &=& \sum_{\bf k,q}\int d{\bf x}\ \frac{1}{2}
im\epsilon_{ij}\left(x_le^{i{\bf k\cdot x}}\right)\left(x_ne^{i{\bf q\cdot
x}}\right)\times\nonumber\\
&&\left\{\begin{array}{l}
\displaystyle\frac{imk_i}{2^{3/2}k\sqrt{\omega_k}}\left[a({\bf
k})+a^\dagger(-{\bf k})\right]+\\[12pt]
\displaystyle\frac{\sqrt{\omega_k}\epsilon_{ir}k_r}{2^{3/2}k}\left[a({\bf
k})-a^\dagger(-{\bf k})\right]\end{array}\right\}
\left\{\begin{array}{l}
\displaystyle\frac{imq_j}{2^{3/2}q\sqrt{\omega_q}}\left[a({\bf
q})+a^\dagger(-{\bf q})\right]+\\[12pt]
\displaystyle\frac{\sqrt{\omega_q}\epsilon_{js}q_s}{2^{3/2}q}\left[a({\bf
q})-a^\dagger(-{\bf q})\right]\end{array}\right\},
\end{eqnarray}
and from $x_le^{i{\bf k\cdot x}} = -i\displaystyle\frac{\partial}{\partial
k_l}\,e^{i{\bf k\cdot x}}$ and $x_ne^{i{\bf q\cdot x}} =
-i\displaystyle\frac{\partial}{\partial q_n}\,e^{i{\bf q\cdot x}}$, and
integrating by parts,
\begin{eqnarray}
X_{ln} &=& -\sum_{\bf k,q}\int d{\bf x}\ \frac{1}{2} im\epsilon_{ij}e^{i{\bf
(k+q)\cdot x}}\times\nonumber\\
&&\frac{\partial}{\partial k_l}\left\{\begin{array}{l}
\displaystyle\frac{imk_i}{2^{3/2}k\sqrt{\omega_k}}\left[a({\bf
k})+a^\dagger(-{\bf k})\right]+\\[12pt]
\displaystyle\frac{\sqrt{\omega_k}\epsilon_{ir}k_r}{2^{3/2}k}\left[a({\bf
k})-a^\dagger(-{\bf k})\right]\end{array}\right\}
\frac{\partial}{\partial q_n}\left\{\begin{array}{l}
\displaystyle\frac{imq_j}{2^{3/2}q\sqrt{\omega_q}}\left[a({\bf
q})+a^\dagger(-{\bf q})\right]+\\[12pt]
\displaystyle\frac{\sqrt{\omega_q}\epsilon_{js}q_s}{2^{3/2}q}\left[a({\bf
q})-a^\dagger(-{\bf q})\right]\end{array}\right\}.
\label{eq:xln}
\end{eqnarray}
After extensive manipulations, almost all contributions to Eq.~(\ref{eq:xln})
cancel, leaving, however, a single residue stemming from the $\delta({\bf k})$
that originates from Eq.~(\ref{eq:result2}):
\begin{equation}
X_{ln} = \frac{i\pi m^2\epsilon_{ln}}{16\pi}\,a^\dagger({\bf 0})a({\bf 0}).
\end{equation}
The existence of this residue is in conflict with Eq.~(\ref{eq:xln0}).

\end{document}